\documentclass[sn-mathphys,Numbered]{sn-jnl}% Math and Physical Sciences Reference Style
\usepackage{soul}
\usepackage{graphicx}%
\usepackage{multirow}%
\usepackage{amsmath,amssymb,amsfonts}%
\usepackage{amsthm}%
\usepackage{mathrsfs}%
\usepackage[title]{appendix}%
\usepackage{xcolor}%
\usepackage{textcomp}%
\usepackage{manyfoot}%
\usepackage{booktabs}%
\usepackage{algorithm}%
\usepackage{algorithmicx}%
\usepackage{algpseudocode}%
\usepackage{listings}%
\usepackage{CJK}
\usepackage{dcolumn}   % needed for some tables
\usepackage{amssymb}   % for math
\usepackage[utf8]{inputenc}
\usepackage{lscape}
\usepackage{booktabs,longtable}
\usepackage{mathptmx, courier, pifont}
\usepackage[scaled=0.92]{helvet}
\usepackage[T1]{fontenc}
\usepackage{textcomp}
\usepackage{rotating}
\usepackage{verbatim}
\usepackage{picinpar}
\usepackage{wrapfig}
\usepackage{float}
\usepackage{multirow}
 \usepackage{lineno}
\usepackage[normalem]{ulem}
\usepackage{flushend}
\theoremstyle{thmstyleone}%
\theoremstyle{thmstyletwo}%
\theoremstyle{thmstylethree}%
\begin{document}
% \linenumbers

%\title[Article Title]{First evidence for pygmy dipole resonance built on excited states in neutron-rich nuclei \textcolor{red}{with non-1$^-$ spin-parity}}
%\title[Article Title]{\textcolor{blue}{First evidence for the  components of the pygmy dipole resonance in a neutron-rich nucleus}}
\title[Article Title]{First evidence for the J$>$1 
%multipole 
% non-dipole(?)
components of the pygmy dipole resonance in neutron-rich nuclei}
%\author*[1,2]{\fnm{R.} \sur{Li}}\email{ren.li@kuleuven.be}
\author*[1,2]{\fnm{R.} \sur{Li}}\email{liren824@gmail.com}
\author[3,4,5]{\fnm{E.} \sur{Litvinova}}
\author[6]{\fnm{M. N.} \sur{Harakeh}}
\author[1]{\fnm{D. } \sur{Verney}}
\author[1]{\fnm{I. } \sur{Matea}}

\author[1,7]{\fnm{L. Al } \sur{Ayoubi}}
\author[8]{\fnm{H. } \sur{Al Falou} }
\author[9]{\fnm{P.} \sur{Bednarczyk}}
\author[10]{\fnm{G. } \sur{Benzoni}}
\author[11]{\fnm{V. } \sur{Bozkurt} }
\author[10,12]{\fnm{A. } \sur{Bracco} }
\author[9]{\fnm{M. } \sur{Ciema\l{}a}}
\author[10,12]{\fnm{F. C. L. } \sur{Crespi}}
\author[1]{\fnm{I.} \sur{Deloncle} }
\author[13]{\fnm{S.} \sur{Ebata} }
\author[14]{\fnm{A.} \sur{Gottardo}}
\author[15]{\fnm{K. } \sur{Hadyńska-Klęk} }
\author[1,16]{\fnm{N. } \sur{Jovancevic} }
\author[7]{\fnm{A. } \sur{Kankainen}}
\author[9]{\fnm{M. } \sur{Kmiecik}}
\author[9]{\fnm{A. } \sur{Maj}}
\author[17]{\fnm{T. } \sur{Mart\'{i}nez}}
\author[18]{\fnm{V.} \sur{Nanal}}
\author[19]{\fnm{O. } \sur{Stezowski}}

\affil[1]{\orgdiv{Université Paris-Saclay}, \orgname{CNRS/IN2P3, IJCLab}, \orgaddress{\city{Orsay}, \postcode{91405}, \country{France}}}
\affil[2]{\orgdiv{Institute for Nuclear and Radiation Physics}, \orgname{KU Leuven}, \orgaddress{\city{Leuven}, \postcode{B-3001}, \country{Belgium}}}
\affil[3]{\orgdiv{Department of Physics}, \orgname{Western Michigan University}, \orgaddress{\city{Kalamazoo}, \postcode{MI 49008}, \country{USA}}}
\affil[4]{\orgdiv{Facility for Rare Isotope Beams}, \orgname{Michigan State University}, \orgaddress{\city{East Lansing}, \postcode{MI 48824}, \country{USA}}}
\affil[5]{\orgdiv{GANIL}, \orgname{CEA/DRF-CNRS/IN2P3}, \orgaddress{\city{F-14076 Caen}, \country{France}}}
\affil[6]{\orgdiv{ESRIG}, \orgname{University of Groningen}, \orgaddress{\street{Zernikelaan 25}, \city{9747 AA Groningen}, \country{The Netherlands}}}
\affil[7]{\orgdiv{Department of Physics, Accelerator Laboratory}, \orgname{University of Jyväskylä}, \orgaddress{\street{P.O. Box 35}, \city{Jyväskylä}, \postcode{FI-40014}, \country{Finland}}}
\affil[8]{\orgdiv{Faculty of Sciences 3}, \orgname{Lebanese University}, \city{Michel Slayman Tripoli Campus, Ras Maska 1352}, \country{Lebanon}}
\affil[9]{\orgdiv{Institute of Nuclear Physics}, \orgname{Polish Academy of Sciences}, \city{Krakow}, \country{Poland}}
\affil[10]{\orgdiv{INFN, sezione di Milano}, \orgname{Dipartimento di Fisica}, \city{Milano}, \country{Italy}}
\affil[11]{\orgdiv{Science Faculty, Department of Physics}, \orgname{Nigde University}, \city{Nigde}, \country{Turkey}}
\affil[12]{\orgdiv{Department of Physics}, \orgname{University of Milan}, \city{Milan, I-20133}, \country{Italy}}
\affil[13]{\orgdiv{Graduate School of Science and Engineering}, \orgname{Saitama University}, \city{Saitama}, \country{Japan}}
\affil[14]{\orgdiv{Laboratori Nazionali di Legnaro}, \city{I-35020 Legnaro}, \country{Italy}}
\affil[15]{\orgdiv{Heavy Ion Laboratory}, \orgname{University of Warsaw}, \country{Poland}}
\affil[16]{\orgdiv{Faculty of Science}, \orgname{University of Novi Sad}, \city{Novi Sad}, \country{Serbia}}
\affil[17]{\orgdiv{Centro de Investigaciones Energ{\'e}ticas}, \orgname{Medioambientales y Tecnol{\'o}gicas (CIEMAT)}, \city{Madrid}, \country{Spain}}
\affil[18]{\orgdiv{Department of Nuclear and Atomic Physics}, \orgname{Tata Institute of Fundamental Research}, \city{Mumbai - 400005}, \country{India}}
\affil[19]{\orgdiv{Universite Claude Bernard Lyon 1}, \orgname{CNRS/IN2P3, IP2I Lyon, UMR 5822}, \city{Villeurbanne, F-69100}, \country{France}}

\date{\today}
\abstract{
Gamma ($\gamma$) decay shapes the synthesis of heavy elements in neutron-rich nuclear environments of neutron star mergers,  supplying the Universe with heavy elements.
The low-energy pygmy dipole resonance (PDR) influences nuclear reaction rates of the rapid nucleosynthesis through enhanced $\gamma$ transitions. However, since it is difficult to reproduce astrophysical conditions in laboratories, PDR was previously observed only in $J = 1$ spin states. Here we report the first experimental observation of $J > 1$ components of PDR, identified in the $\beta$-delayed $\gamma$ decay of the J$^{\pi}$ = 3$^{-}$ spin-parity isomer of $^{80}$Ga.
%Gamma decay plays a crucial role in the radioactivity of newly synthesized neutron-rich atomic nuclei in neutron star mergers,%contributing to their energy output heating the ejecta. 
%Such nuclei embedded in astrophysical plasma undergo various reactions, such as beta decay, neutron capture, and gamma 
%transitions between nuclear excited states
%emission and absorption, which determine the rate and path of nucleosynthesis. The knowledge about the nuclear low-energy dipole %gamma transitions, associated with the pygmy dipole resonance (PDR), is critical for understanding this process. 
%Since it is difficult to reproduce astrophysical conditions in laboratories, only gamma decays from the J$^{\pi}$ = 1$^{-}$ spin-parity states %have so far been firmly assigned to PDR. However, its higher-multipole components are needed for the complete and accurate modeling %of stellar nucleosynthesis. We measured such components in the form of beta-delayed gamma rays in the energy window [0 - 10.312(4) %MeV] opened in the beta decay of the J$^{\pi}$ = 3$^{-}$ isomeric state of gallium $^{80}$Ga. %$^{80g+m}$Ga. 
The data analysis, combined with decay information and theoretical calculations allows the identification of resonant structures below the neutron emission threshold of the neutron-rich germanium $^{80}$Ge as J$^{\pi} = (2,3)^-$ components of the PDR built on the low-lying J$^{\pi}$ = 2$^+$ quadrupole state.  
%The results are confronted with the existing knowledge about the microscopic structure of the regular single-component J$^{\pi}$ = 1$^{-}$ PDR, suggesting that %higher-multipole components of the PDR built on a quadrupole state are identified. 
Our findings extend the concept of PDR beyond dipole states, with implications for nuclear structure theory and experiment, as well as the element production in the cosmos.}
%}
%have far-reaching implications for nuclear structure and astrophysics, opening an avenue for experimental and theoretical studies of the %multicomponent PDR and its impact on element production in the cosmos.

%\pacs{}

\maketitle

\section*{Astrophysical relevance of pygmy dipole resonance}
\begin{figure*}[!htb]
\vspace{-1cm}
\centering
\includegraphics[width=1.1\textwidth]{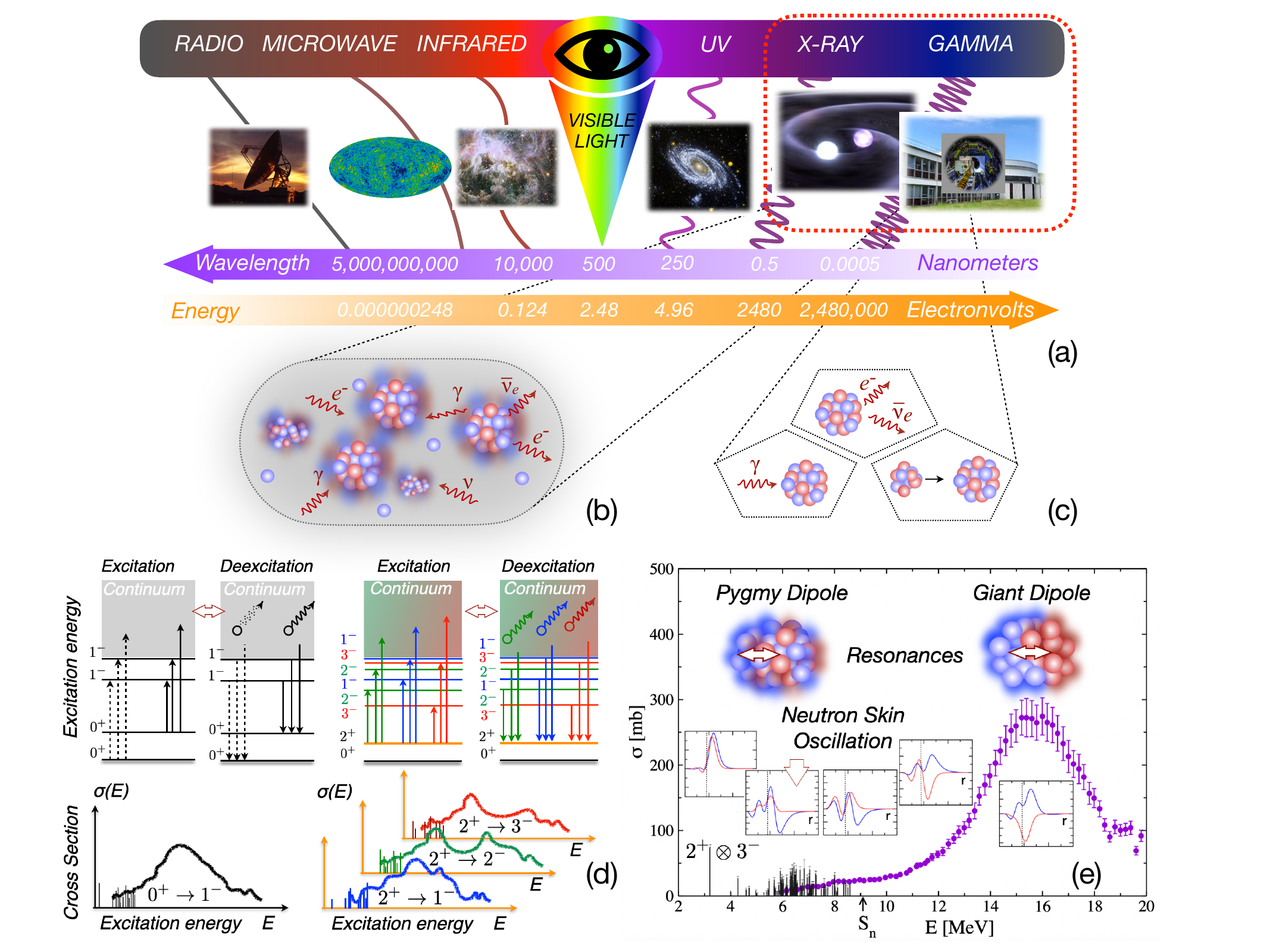}
\caption{(a) Characteristic scales of the electromagnetic spectrum of space, astronomical objects, and laboratory nuclear astrophysics \cite{nasa}; (b) Schematic illustration of atomic nuclei embedded in a stellar environment vs (c) selective probing nuclei in a laboratory; (d) Schematic spectra of transitions between the nuclear energy levels: E1 excitations from J$^{\pi}$ = 0$^{+}$ ground or excited states, only possible to J$^{\pi}$ = 1$^{-}$ levels (left), vs E1 excitations from J$^{\pi}$ = 2$^{+}$ excited states, ending in a multiplet of J$^{\pi}$ = \{1,2,3\}$^{-}$ levels (right); (e) A typical dipole spectrum of a medium-mass neutron-rich nucleus (stable $^{120}$Sn) dominated by the low-energy pygmy and high-energy giant resonances, from left to right: the lowest two-phonon $[2^+\otimes3^-]^{1-}$ state and the discrete dipole states pertaining to the lower part of PDR and observed in the gamma ($\gamma,\gamma'$) scattering experiments (black bars) \cite{OezelTashenov2014}, 
overlapping with the data on the proton ($p,p'$) scattering (purple error bars) \cite{Bassauer2020} which span both the PDR and GDR energy domains. The insets show the proton (red) and neutron (blue) transition densities characterizing the E1 transitions from the ground to J$^{\pi} = 1^{-}$ states in the respective energy intervals as functions of the radial distance $r$ from the nuclear center. The vertical lines mark the nuclear radius.}
\label{fig0}
\end{figure*}

The most violent events in the Universe, such as neutron star mergers (NSMs), pulsars, supernova explosions, and black hole dynamics, are natural laboratories of the extreme states of matter, light, and their interaction. A large portion of information from such events as NSMs studied by multimessenger astrophysics is communicated to us by gamma ($\gamma$) rays from the short-wavelength part of the spectrum illustrated in Fig. \ref{fig0}(a). 
%Modern nuclear astrophysics suggests that 
NSMs are the primary astrophysical sites of nucleosynthesis, where
%, forming heavy atomic nuclei. 
beta ($\beta$) decay and rapid radiative neutron capture ($n$,$\gamma$) are the two alternating phases, producing
% of this process, are established as the prevailing mechanism for synthesizing 
heavy chemical elements beyond iron in our Universe \cite{Pian2017}. Atomic nuclei embedded in stellar environments, schematically depicted in Fig. \ref{fig0}(b), undergo 
%specific 
reactions caused by weak, strong, and electromagnetic forces.
Gamma decay contributes significantly to the cooling of the neutron-rich matter, formed and ejected from NSMs, and plays the role of an essential messenger, alongside neutrinos and gravitational waves, in multimessenger astrophysics \cite{Meszaros2019}. Gamma emission and absorption by atomic nuclei, along with other reactions, are actively studied in experimental laboratories under terrestrial conditions, symbolized by Fig. \ref{fig0}(c). 
%{\textcolor{red}{as limited by a single probe and initial nuclear ground state}. 
On the theory side, the effort is directed toward maximizing the predictive power of extrapolations to the limits of nuclear stability as well as stellar temperatures and densities for providing inputs for modeling and understanding cataclysmic events in the cosmos.

In the context of stellar nucleosynthesis, the pygmy dipole resonance (PDR) is among the most relevant 
%{\textcolor{red}{
nuclear vibrational modes \cite{bracco2019isoscalar}. It has originally been reported as a new hypothetically collective excitation located on the low-energy tail of the isovector giant dipole resonance (IVGDR), which is a broad peak
in the electric dipole (E1) $\gamma$-ray transition probability distribution, or strength,
occurring due to the out-of-phase oscillation of the proton and neutron Fermi liquids against each other and centered at 15-20 MeV \cite{harakeh2001giant}. 
The PDR enhancement of the E1 strength takes place around 5-7 MeV \cite{bartholomew1961neutron} in heavy nuclei, near and below the neutron separation threshold (S$_n$), and can be partly above S$_n$ in 
%lighter 
neutron-rich nuclear species \cite{rossi2013measurement}. PDR is more pronounced in neutron-rich nuclei and has thus been interpreted as an oscillation of the neutron excess against a proton-neutron saturated core \cite{mohan1971three,Vretenar2001}. Recently, PDR has been shown to have a strongly mixed isoscalar and isovector character \cite{bracco2019isoscalar, savran2013experimental}, linked to the neutron-skin thickness \cite{piekarewicz2011pygmy},
%reinhard2010information,}, 
and related, 
%\textcolor{red}{
via the electric dipole polarizability, to the slope parameter of the symmetry energy in the nuclear matter equation of state \cite{klimkiewicz2007nuclear,Roca2015}. Thereby, nuclear PDR is an essential benchmark for modeling dense astrophysical objects, such as neutron stars \cite{fattoyev2018neutron}, 
%or quark stars \cite{bombaci2021gw190814}, 
in particular, quantifying their mass-radius relation. The PDR's distribution around S$_n$
% is the doorway for the neutron capture, 
is critical for determining the nucleosynthesis path, in which the ($\beta$, $\gamma$) reaction competes with the beta-delayed neutron emission ($\beta$,$n$). In the photon bath of the NSM environment, PDR above S$_n$ enhances the reverse process of ($\gamma$,$n$) photodisintegration. Therefore, the accurate knowledge of PDR in neutron-rich nuclei is critical for understanding the production of heavy elements beyond iron in the Universe \cite{goriely1998radiative}.

\section*{Structural properties and identification of the pygmy dipole resonance }

The underlying structure of PDR is under active debate and scrutiny. It was found to be more sensitive to the detailed properties of the nuclear forces than IVGDR \cite{savran2013experimental,bracco2019isoscalar,spieker2020accessing,paar2007exotic}, which stipulated the use of a variety of experimental probes, from electromagnetic, such as bremsstrahlung, Compton backscattering and Coulomb excitation, to hadronic using proton or $\alpha$ scattering, to weak employing $\beta$-decay. The 
%\textcolor{red}{
PDR's isospin splitting was revealed in complementary experiments with isoscalar ($\alpha$, ${\alpha}^{\prime}$) and isovector ($\gamma$,${\gamma}^{\prime}$) probes of nuclei $^{138}$Ba \cite{endres2009splitting}, $^{140}$Ce \cite{savran2006nature}, and $^{124}$Sn \cite{endres2010isospin},
%. Microscopic theory interpretations 
suggesting that PDR splits into a lower-energy isoscalar surface mode and a higher-energy component with an isovector admixture from IVGDR  \cite{paar2009isoscalar,endres2010isospin,lanza2014dipole}. Another milestone was the identification of a two-peak PDR structure with a single probe in $^{142}$Nd \cite{Angell2012}, $^{60}$Ni \cite{scheck2013decay}, $^{130}$Te \cite{isaak2013constraining} and $^{140}$Ce \cite{loher2016decay} using fully linearly polarized, quasi-monochromatic, Compton-backscattered photons, 
%which suggested 
indicating that 
%\textcolor{red}{
the lower peak preferably decays directly to the ground state, while the upper part of PDR cascades via excited states,
%\sim$25\% 
%This is consistent 
in agreement with the recent systematic studies of 
%the PDR splitting in 
the tin isotopic chain \cite{Markova2025}. 
%}
%showing that the lower peak is dominated by neutron oscillations and preferably decays to the ground state, while the upper one has a larger proton %contribution and is, therefore, more likely to cascade via excited states. 
Theoretical analyses allow one to differentiate PDR from IVGDR via the neutron and proton transition densities characterizing the wave functions beyond one-particle-one-hole ($1p1h$) configurations with respect to the Fermi energy. An example of one of the most studied stable neutron-rich nuclei $^{120}$Sn is displayed in Fig. {\ref{fig0}}(e). 
Along with the experimental spectrum of E1 states, it shows characteristic proton and neutron transition densities for different spectral energy regions, obtained within the relativistic equation of motion (REOM) method  \cite{PhysRevC.100.064320,novak2024} detailed in the section "Methods: Theoretical approach".
The lowest-energy dipole excitations exhibit nearly pure in-phase, or isoscalar, behavior of the proton and neutron transition densities; a pronounced neutron dominance appears beyond the surface in the states concentrated 
%in the first PDR peak 
around $\sim$6.5 MeV; then the proton contribution increases 
%in the second PDR 
at the peak around $\sim$8 MeV, where the transition densities start also showing some decoherence,
% in the nuclear bulk, 
which is attributed to the IVGDR admixture. With further energy increase, the decoherence increasingly takes over until the proton and neutron components become completely out of phase at the IVGDR peak. 
% mention here LEDS and PDR
Similar patterns were identified in other nuclei and thus adopted to distinguish PDR from IVGDR, as well as to differentiate groups of states within PDR.

\section*{PDR built on excited states and its non-dipole components}

%\textcolor{blue} {
Both the ground states of even-even nuclei (i.e., nuclei with even numbers of protons and neutrons) with $J^{\pi} = 0^{+}$ and their low-energy excited states with $J^{\pi} = (0,1,2)^{+}$
%both $J^{\pi} = 0^{+}$ and $J^{\pi} = 2^{+}$ 
can be the final states of E1 de-excitations of $J^{\pi} = 1^{-}$ PDR states, by virtue of angular momentum algebra. The cases of $J^{\pi} = (0,2)^{+}$ were considered, e.g., in Refs. \cite{Tonchev2010,Angell2012,Derya2013,mashtakov2021structure} with the analyses of the corresponding branchings.  
In the early stages of the rapid nucleosynthesis, known as r-process, the neutron capture $(n, \gamma)$ and neutron emission $(\gamma, n)$ are in equilibrium. Due to the finite temperature of the stellar environment, atomic nuclei in the plasma are in their excited states, which means that excitations and de-excitations between the lowest-energy excited states and higher-lying PDR states occur, and both impact the r-process scenario. Probabilities of such transitions are related to each other via the detailed balance \cite{bohr1998nuclear}. In turn, E1 excitations from an initial ($i$) state with spin-parity $J_i^{\pi_i}$, where $J_i  > 0$, end up in the triplet of final ($f$) states with $J_f^{\pi_f} = (J_i-1, J_i, J_i+1)^{-\pi_i}$ over the PDR and IVGDR energy domains, but for $J_i^{\pi_i}  = 0^+$ only $J_f^{\pi_f} = 1^-$ final states are allowed by the angular momentum and parity conservation, as illustrated in Fig. \ref{fig0}(d). While the majority of the reported PDR studies have been focused on de-excitations of $J^{\pi} = 1^{-}$ PDR states, the PDR components with higher angular momenta remained largely unknown. 
% Hence, one can make the assumption that PDR based on excited states can exist which can be built on 2$^{+}$ or 4$^{+}$ levels, similar to IVGDR built on low-%lying discrete levels \cite{snover1986giant,harakeh2001giant} and similar also to recent studies \cite{martin2017test,markova2021comprehensive} which present %evidences for supporting the validity of Brink-Axel hypothesis (BAH) \cite{brink1955some,axel1962electric,von2019electric} in PDR energy region. 
%}

%\textcolor{blue} {
Furthermore, besides few exceptions \cite{Adrich2005,wieland2009search,rossi2013measurement,Wieland2018}, the PDR studies are mostly restricted to stable or near-stable nuclei \cite{zilges2022photonuclear,tamii2011complete,savran2006nature,crespi2014isospin,weinert2021microscopic},
%,von2019electric} 
because of experimental limitations, such as the low intensity of secondary beams and the low detection efficiency of 
%high-energy 
$\gamma$ rays in conventional experiments with exotic nuclei, although neutron-rich unstable medium-heavy nuclei dominate the r-process path. 
To overcome these limitations, beta decay from ground and isomeric states of odd-odd neutron-rich nuclei, whose spin-parities are variable, can be used for the selective population of the
%$J > 1$ 
PDR states in their residual, or "daughter", even-even nuclear systems. 
Thanks to the large Q$_{\beta}$ values in neutron-rich nuclei, considerable fractions of PDR can be thereby reached and studied via observation of their subsequent gamma decays. Pioneering experiments using $\beta$ decay were performed in Ref. \cite{scheck2016investigating}, followed by Refs. \cite{gottardo2017unexpected} and \cite{mashtakov2021structure}, where the strength distributions extracted from E1 $\gamma$ de-excitations of the $\beta$-populated states were confronted with those obtained in ($\gamma$, $\gamma^{\prime}$) reactions. The method showed its effectiveness
% in investigating the underlying structure and decay pattern of the 
for $J = 1$ PDR states and thus admits an extension to the unknown non-dipole, or $J>1$, PDR components. This opportunity was realized experimentally and theoretically in our work.
%, as detailed in the following sections.
%}
 
%The counterpart of these advantages is that such measurements are still challenging due to Pandemonium effect \cite{hardy1977essential}.

\section*{Spectroscopy of the $\beta$-delayed $\gamma$ emission}

\begin{figure*}[!htb]
%\vspace{-1cm}
\centering
%\hspace{-1.5cm}
\includegraphics[width=0.90\textwidth]{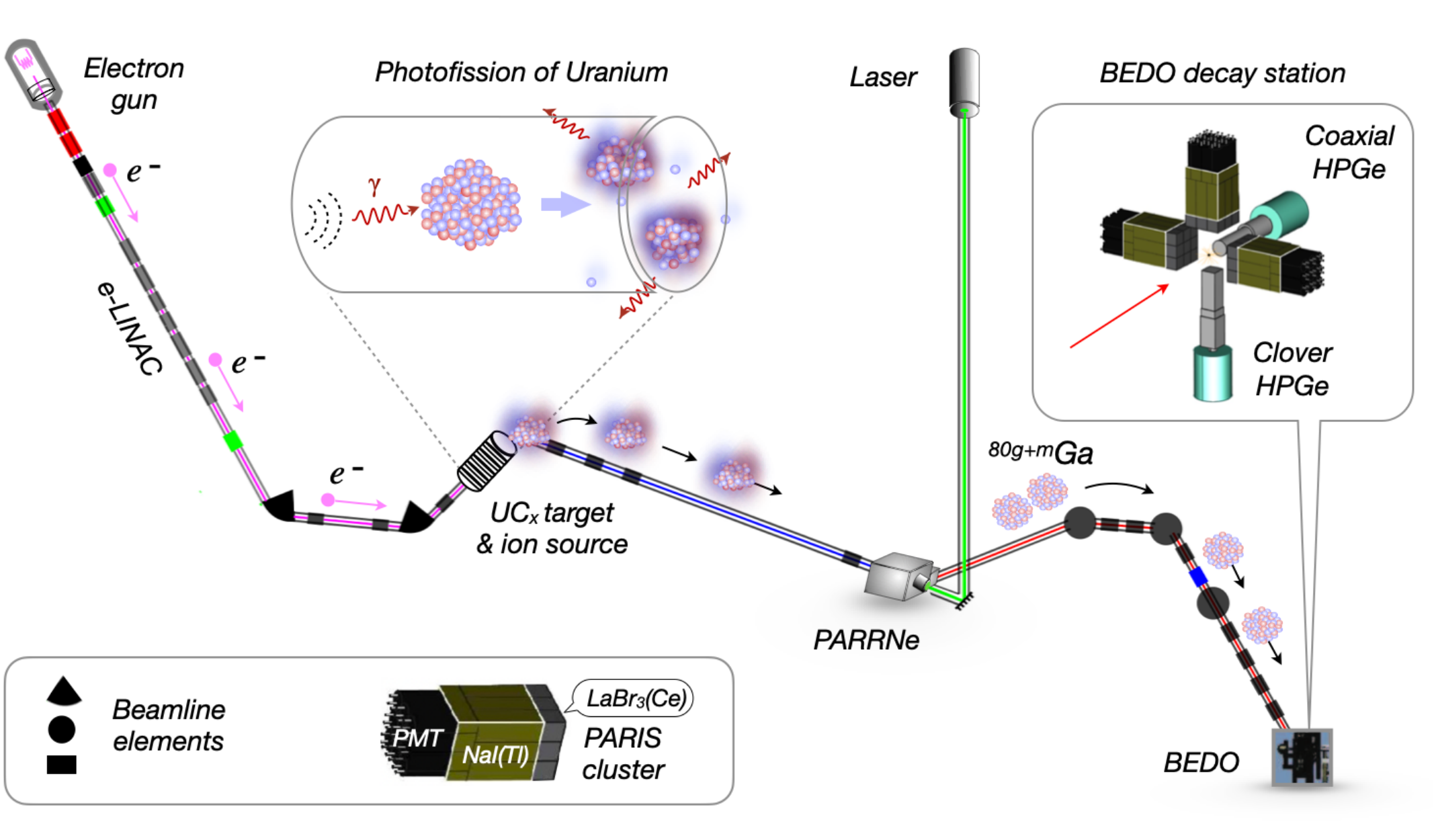}
\vspace{0.5cm}
\caption{
%\textcolor{blue}{
Schematic design of the ALTO ISOL facility (see text for details). 
%In this experiment, a $^{80g+m}$Ga beam was produced by the photo-fission of a UC$_x$ target induced by a 50-MeV electron beam with an intensity of ~7 $\mu$A. For the beam purification, the laser-ionized Ga beam was mass-selected by PARRNe (Production d’Atomes Radioactifs Riches en Neutrons). Then, a pure $^{80g+m}$Ga beam was directed to and collected by the movable tape system of decay station BEDO. The $^{80g+m}$Ga beam intensity is of $\approx$10$^4$ pps. 
The inset displays the scheme of BEDO with two HPGe detectors for the low-energy $\gamma$-ray detection with high energy resolution and three PARIS clusters for high-energy $\gamma$-ray detection with high efficiency, mounted with a cylindrical plastic detector for $\beta$-tagging (not shown).
%}
}
\label{fig1_0}
\end{figure*}

%In this letter, 
%\textcolor{blue}{
Here, we report on the first evidence of the population of $J>1$ PDR states
%built on low excited states 
by the $\beta$-decay of $^{80}$Ga. 
This endeavor was triggered by an unexpectedly strong competition between neutron and $\gamma$ emission from excited states of $^{83}$Ge populated by $\beta$-decay of $^{83}$Ga \cite{gottardo2017unexpected}. 
%Indeed, combined with the Gamow-Teller transition probability (B(GT)) measurements, the spectroscopic data (energies, spins, and parities) of the states provide %clues on the PDR structure, which are important observables to check the calculated wave functions from theoretical models.
The measurement was performed at the Accélérateur Linéaire et Tandem à Orsay (ALTO) ISOL facility \cite{ibrahim2007alto}, illustrated schematically in Fig. \ref{fig1_0}. A radioactive low-energy ion beam of $^{80}$Ga, containing the long-lived isomer $^{80m}$Ga and the ground-state $^{80g}$Ga (both are regarded as "isomers" below) with J$^{\pi}_{m}$ = 3$^{-}$ and J$^{\pi}_{g}$ = 6$^{-}$, respectively \cite{cheal2010discovery}, 
%Q$_{\beta}$ = 10.312(4) MeV \cite{wang2017ame2016}, 
was produced by the photo-fission of a uranium carbide (UC$_x$) target induced by a 50-MeV electron beam with an intensity of ~7 $\mu$A from a linear electron accelerator (e-LINAC). For the beam purification, the laser-ionized Ga beam was mass-selected by the PARRNe (Production d’Atomes Radioactifs Riches en Neutrons) magnetic separator. Since around the nuclear mass number $A = 80$ the only surface-ionized component of a photo-fission generated ion beam is Ga, a complete isotope purity was achieved. Then, the $^{80g+m}$Ga beam with the yield of $\sim$10$^{4}$ pps was 
%directed and 
collected by the tape system of the BEDO (BEta Decay studies at Orsay) decay station, which was moved periodically to minimize the activity of the daughter nucleus \cite{etile2015low}. 
%The time settings were 0.5 s for the background, 5 s for the ion collection, and 5 s for the decay measurement.
%}

\begin{figure*}[!htb]
\centering
\includegraphics[width=1.0\textwidth]{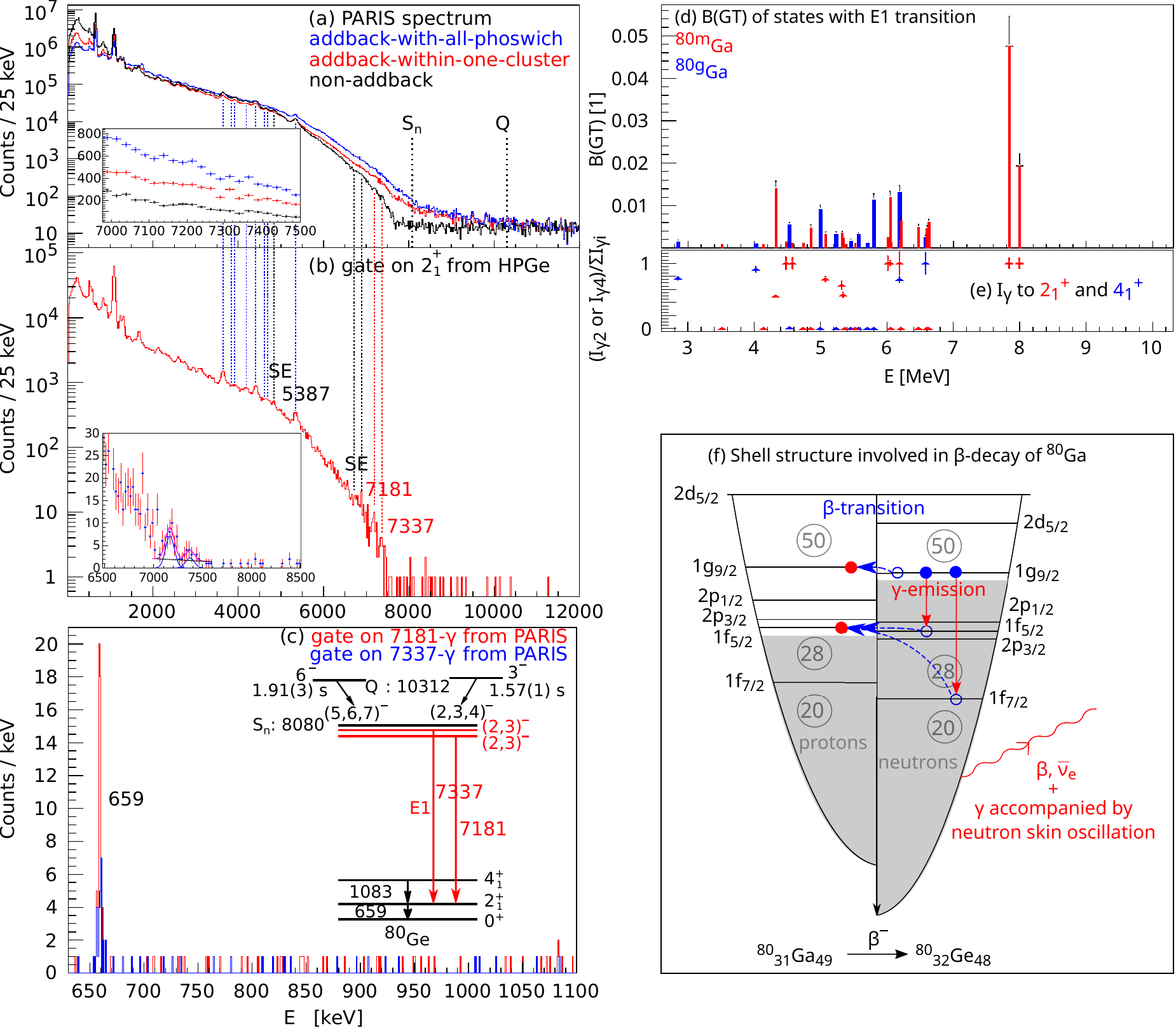}
\caption{(a): High-energy $\beta$-gated $\gamma$-ray spectrum from PARIS, black in mode 1, red in mode 2, and blue in mode 3. 
Inset: spectrum 
%with 
% \textcolor{blue}{
on a linear scale for the energy region from 7.0 to 7.5 MeV; (b): $\gamma$-rays of the spectrum shown in red in panel (a), in coincidence with the 659 keV $\gamma$-ray. The $\gamma$-rays with 
%strong 
%\textcolor{blue}{
E1 multipolarities are marked with vertical dashed lines at 3664.6, 3818.6, 3920.4, 4225.1, 4412.8, 4665.3, 4678.5, 5387.4, 7181, and 7337 keV. Inset: 
spectrum on a linear scale for the energy region from 6.5 to 8.5 MeV and the fitting curves; (c): The inverse coincidence: $\gamma$-rays registered in HPGe in coincidence with the 7181 keV (red color) and 7337 keV (blue color) $\gamma$-rays from PARIS, 
%\textcolor{blue}{
gated on the red histogram of panel (a). Inset: simplified level scheme showing the E1 
%\textcolor{blue}{
transitions from the excited states at 7840 keV and 7996 keV, located below S$_n$ and identified with $J^{\pi} = (2,3)^-$, to the lowest 2$_1^+$ state; (d): The measured B(GT) distribution of $^{80m,g}$Ga versus the excitation energy of 
%\textcolor{blue}{
the states in $^{80}$Ge decaying by E1 transitions; (e): The decay branching ratio to 2$_1^+$ and 4$_1^+$ for 
%\textcolor{blue}{
the states from panel (d); (f): Schematic representation of the shell structure involved in the $^{80m,g}$Ga $\beta$-decay, wherein particles (holes) are shown as filled (open) circles, and $\beta$ and $\gamma$ transitions of unspecified complexity are associated with blue and red arrows, respectively.} 
%The decays to the $1p1h$ and $2p2h$ configurations are indicated by blue arrows}. The single-particle levels %in $^{80}$Ge 
%are \textcolor{blue}{taken from the Hartree-Fock-Bogoliubov calculations}. }
\label{fig1}
\end{figure*}

%\textcolor{blue}{
The radiation emitted was detected by a hybrid array around the collection spot. The array consisted of one cylindrical plastic scintillator for $\beta$-tagging with the efficiency of $\sim$70$\%$, surrounded by two high-purity germanium detectors (HPGe) and three PARIS (Photon Array for studies with Radioactive Ion and Stable beams) clusters \cite{ciemala2009measurements,ghosh2016characterization} comprising twenty-seven optically isolated segmented phoswiches. Each phoswich was composed of two layers of different scintillators and a photomultiplier (PMT): a 2"$\times$2"$\times$2" Lanthanum Bromide/Cerium Bromide (LaBr$_3$/CeBr$_3$) crystal and a 2"$\times$2"$\times$6" Thallium-doped Sodium Iodide (NaI(Tl)) one, which form an assembly sensitive to a wide range of $\gamma$-ray energies. The high energy resolution of HPGe of 2.611 keV 
at 1083.6 keV 
%makes the $\gamma$-$\gamma$ coincidence technique 
%$\gamma$-$\gamma$ coincidence technique is not explained
%very efficient, not only to 
enabled an efficient reconstruction of the $\gamma$ transition cascades by $\gamma$-$\gamma$ coincidences with a considerable background suppression 
%\textcolor{red}
%{
as described in the section "Methods: Multiplicity of high-energy $\gamma$-rays and background assessment".
%} 
Furthermore, PARIS has a high detection efficiency in the Q$_{\beta}$ window of $^{80}$Ga, i.e., the energies below 10.312(4) MeV relative to the ground state of $^{80}$Ge  \cite{wang2017ame2016}. The detectors were energy calibrated up to 9 MeV as detailed in Ref. \cite{ren2022thesis}.
%using sources including $^{58}$Ni(n-th,$\gamma$) with an energy of 8999.267(15) keV.
%
%
The data were acquired in a triggerless mode. For an accurate extraction of the Gamow-Teller transition probability (B(GT)) values, the Compton scattering and single escape (SE) events were suppressed thanks to the specific properties of PARIS: (i) the presence of two shells, where the outside one plays the role of veto detector 
%to perform anti-coincidence operation for
rejecting events, where the $\gamma$-rays did not deposit their full energy in the inner shell, by anti-coincidence operation; (ii) the ability to reject pileup events through 
%the technique of 
pulse-shape analysis \cite{ren2022thesis}. 
Three data sorting modes of PARIS were employed: 1) LaBr$_3$(Ce) considered individually; 2) add-back performed within one cluster; 3) add-back performed within three clusters (full PARIS). For each of these modes, the outer-shell NaI(Tl) crystals work as a vetoing shield. The energy resolution and detection efficiency of PARIS are 112 keV and 0.30(2)$\%$, respectively, at 7.18 MeV under mode 3 \cite{ren2022thesis}.
Fig. \ref{fig1}(a) shows the $\beta$-gated $\gamma$ spectrum measured with PARIS. Above the energy of 5.6 MeV, only $\beta$-delayed $\gamma$ rays from $^{80}$Ga exist since the Q$_{\beta}$ values of the daughter and granddaughter nuclei $^{80}$Ge and $^{80}$As are 2679(4) keV and 5545(4) keV, respectively \cite{wang2017ame2016}. Thus, below we refer to the $\gamma$-rays emitted from the excited states of $^{80}$Ge between 5.6 MeV and the 
endpoint energy as high-energy $\gamma$-rays. 
% %%%It looks to me that the  was already described in the previous paragraph:
%Due to the add-back operation, the single escape (SE) peaks were reduced, and the background was lowered, thanks to the specific properties of PARIS: (i) it has %two shells so that the outside one could play the role of veto detector to perform anti-coincidence operation for rejecting the non-full energy peaks; (ii) it has the %ability to reject pileup events through the technique of pulse-shape analysis \cite{ren2022thesis}. 
The separation between neutrons and $\gamma$ rays was achieved by the time-of-flight measurement, so the $\gamma$ decay information was extracted directly from the spectrum, avoiding Monte-Carlo simulations or theoretical assumptions.
% about the $\gamma$ strength distribution.
%}

Detailed information was extracted from this high-quality dataset, including the level energies, $\beta$ transition half-lifes, B(GT) values, and the $\beta$ decay pattern to perform the spin-parity assignment. While
Ref. \cite{Li2025a} reports on the simultaneous impacts of nuclear shell structure and collectivity on the beta decay of $^{80}$Ga and Ref. \cite{li_hal05035090} on the detailed level schemes and competition between the Gamow-Teller and first-forbidden transitions in $^{80g+m}$Ga beta decay,
here we focus on the subsequent $\gamma$ decay of $^{80}$Ge, which enables identification of the $J>1$ PDR components.
As presented in Fig. $\ref{fig1}$(b), two distinctly resolved full-energy peaks survived in the PARIS spectrum in coincidence with 659 keV $\gamma$ rays corresponding to the $2^{+}\rightarrow 0^{+}$ transition of $^{80}$Ge. These peaks can be unambiguously 
%interpreted as originating from 
attributed to direct transitions from the excited states located at 7840(53$_{stat}$+18$_{sys}$) and 7996(53$_{stat}$+19$_{sys}$) keV to the lowest quadrupole positive-parity (2$_1^{+}$) state, as indicated in the simplified level scheme in Fig. \ref{fig1}(c). 
%\textcolor{red}{
For further verification, the reversed coincidence spectra gated on the 7181 keV (red color) and 7337 keV (blue color) $\gamma$ transitions in the PARIS spectrum (red histogram in Fig. \ref{fig1}(a)) are shown in Fig. $\ref{fig1}$(c).
%\textcolor{red}{, gates on red histogram in Fig. \ref{fig1}(a)}. 
The blue lines in Fig. \ref{fig1}(b) mark the E1 $\gamma$-rays of this experiment, also observed in the HPGe detector.
%populated in 

\section*{Multipolarity of $\gamma$ transitions}

%\textcolor{red}{
The analysis outlined in the section "Methods:
Precursor identification" concludes that the isomeric state $^{80m}$Ga $\beta$-populates predominantly $J^{\pi} = (2,3,4)^-$ levels, while the $\beta$ decay of $^{80g}$Ga ends up in the $J^{\pi} = (5,6,7)^-$ excited states of $^{80}$Ge. The multipolarities of the $\gamma$ de-excitations of the $(2,3,4)^-$ levels to the $2^{+}_1$ one are 
%\textcolor{red}{
constrained by the angular momentum algebra, allowing for 
%E1, M2, E3; E1, M2, E3 and M2, E3, respectively.
%{\bf Here, it should be 
$J^{\pi} = (1-6)^{-}$.
%} 
%\textcolor{blue}{
However, according to the Weisskopf single-particle estimates for transition probabilities, for the same energy of a transition, the probability decreases by $\sim$3 orders of magnitude when the multipolarity increases by one unit, and for the same multipolarity, the probability of an electric transition is $\sim$2-3 orders of magnitude stronger than for a magnetic transition \cite{bohr1998nuclear}.
%\cite{hamilton1975electromagnetic,ajzenberg2013nuclear}. 
%From the Weisskopf estimates, 
%$T_{1/2}(E1) = 6.627A^{-2/3}E_{\gamma}^{-3}\times10^{-15}$ s and $T_{1/2}(M2) = 3.1A^{-2/3}E_{\gamma}^{-5}\times10^{-8}$ s, where $E_{\gamma}$ is in MeV. 
Particularly, 
the half-lives of the 7.84 MeV and 7.996 MeV states are 
%$9.8\times10^{-19}$ s and $9.2\times10^{-19}$ s, respectively, 
$\sim10^{-18}$ s for the E1 transitions and $\sim10^{-13}$ s for the M2 transitions. Since the dipole $\gamma$ de-excitations dominate, the 4$^-$ states preferentially decay to lower-lying levels with closer spins, such as $J = (3,4,5)^{\pm}$ so that $J^{\pi} = 4^-$ can be ruled out from characterization of the two states of interest. Therefore, $J^{\pi} = (2,3)^{-}$ are the most probable spin-parities of the 
7.84 MeV and 7.996 MeV states, decaying to the $2^{+}_1$ state by E1 transitions.
%\textcolor{red}{
From similar estimates it follows that the $J^{\pi} = (5,6,7)^-$ excited states originating from the $\beta$ decay of $^{80g}$Ga predominantly $\gamma$ decay to $J^{\pi} = 4^+$ and higher-$J$ states.
Fig. \ref{fig1}(d) displays the measured B(GT) distribution of all the states of $^{80}$Ge populated by the $^{80m}$Ga and $^{80g}$Ga $\beta$ decays that $\gamma$ decay by E1 transitions, and Fig. \ref{fig1}(e) collects their $\gamma$ decay branching to the 2$_1^+$ and 4$_1^+$ states, respectively. 
%
%} 
% here we should add the description of Fig. 3f.
% here mention that we focus on the 3- -> 2-,3- -> 2+ chain (?)

Thereby, the 7.84 MeV and 7.996 MeV states, which we refer to as {\it resonant} below, together with the weaker-populated states at lower energies, also $\gamma$ decaying to $2^{+}_1$ by E1 transitions, represent the subthreshold $J^{\pi} = (2,3)^{-}$ components of PDR built on the $2^{+}_1$ state, which we name PDR$^{\ast}(2^+_1)$ from now on. 
Fig. \ref{fig1}(f) provides a shell-model interpretation of the observed decay sequence. 
%structure analysis \cite{grawe2004shell}: 
The  $\beta$ decay 
%corresponds to 
includes the conversion of a deeply bound neutron 
%from the $Z = N = 28$ proton-neutron saturated core 
into a proton at the Fermi surface via the ${\nu}f_{7/2}$ $\to$ ${\pi}f_{5/2}$ spin-flip transition, ending in the $\sim8$ MeV excitation in the daughter nucleus. It de-excites with the $\gamma$ emission 
%accompanied by dropping down of a 
involving the surface neutrons, which may trigger a dipole oscillation of the neutron skin via the $1p1h$ or more complex configurations. 
%Since the deeply bound single-particle states are not involved in this process, the high-energy IVGDR is irrelevant in the dipole channel of the $\beta$-delayed $\gamma$ emission.
%it is impossible for this neutron $\beta$-transition triggered oscillation to involve very deep single-particle levels, 
%This is a key difference in comparing IVGDR depicted as oscillation of all protons against all neutrons with PDR as a property of the neutron skin. %Therefore, results from REOM$^3$ and shell model strongly support our assignments of 7840(53$_{stat}$+18$_{sys}$) keV and 7996(53$_{stat}$     %+19$_{sys}$) keV levels to PDR states.

%Experimentally, we have observed E1 strengths connected to 2$_1^+$ state from below the S$_n$ region in $^{80}$Ge. 
To understand the nuclear structure corresponding to these $\gamma$ transitions, microscopic calculations have been undertaken with the REOM method \cite{PhysRevC.100.064320,novak2024}. REOM is based on exact ab-initio many-body theory, while its implementations approximate the nuclear excited-state wave functions by superpositions of correlated particle-hole ($ph$) configurations with respect to the nuclear Fermi energy. This allows for generating states of growing complexity, providing a higher accuracy description of the nuclear spectral properties. Accordingly, we adopt the notation REOM$^n$, where the upper index $"n"$ is a universal complexity index marking the maximal number of correlated $ph$ configurations in the composition of excited states. Further details are provided in the section "Methods: Theoretical approach".
%}

%\section*{Pygmy dipole resonance built on excited states and neutron-skin oscillation}
\section*{Theoretical analysis}

\begin{figure*}
\begin{center}
\vspace{0.3cm}
\includegraphics[scale=0.35]{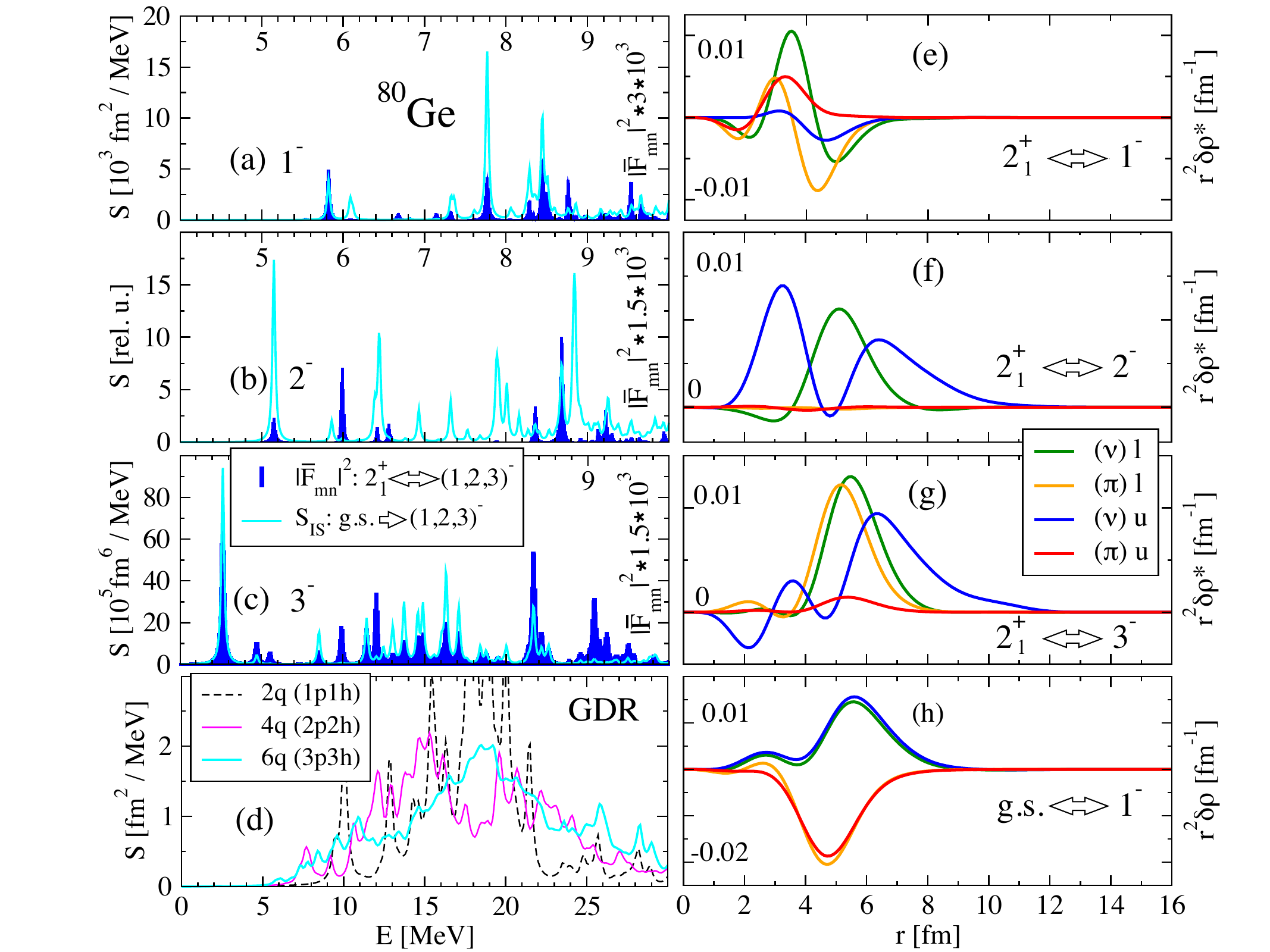}
\end{center}
%\vspace{-6.3cm}
\caption{Left: REOM$^3$ strength $S(E)$ in $^{80}$Ge for the isoscalar (IS) electric dipole $1^-$ (a), magnetic quadrupole $2^-$ (spin-dipole) (b) and electric octupole  $3^-$ (c) transitions from the ground state (g.s.) (solid light blue curves) with the 
%reduced transition probabilities $B_{mn}(E1) = 
squared reduced matrix element $|{\bar F}_{mn}|^2$ [e$^2$ fm$^2$] of E1 transitions between the obtained excited states $(1,2,3)^-$ and the lowest quadrupole state $2_1^+$ (dark blue histograms). Panel (d) displays the IVGDR computed in REOM$^{1-3}$ with {\it correlated} $2q$, $4q$, and $6q$ configurations, respectively. Right: the corresponding neutron ($\nu$) and proton (${\pi}$) transition densities of the characteristic states from the lower (l) and upper (u) parts of the (e,f,g) PDR$^{\ast}(2^+)$ ($\delta\rho^{mn}(1^-) \equiv \delta\rho^{\ast}$); (l): 8.0 -- 9.0 MeV, (u): 9.0 -- 10.0 MeV and (h) IVGDR ($\delta\rho^{0n}(1^-) \equiv \delta\rho$); (l): 17.5 -- 18.5 MeV, (u): 18.5 -- 19.5 MeV.}
\label{theory}%
\end{figure*}

 Fig. \ref{theory} (a-c) displays the REOM$^3$ isoscalar (IS) strength in $^{80}$Ge $S(E) = \sum_{n} B_{0n}\delta_{\Delta}(E-E_n)$ for the transitions from the ground state $|0\rangle$ to $|n\rangle = (1,2,3)^-$ excited states below 10 MeV along with the 
squared reduced matrix elements  $|{\bar F}_{mn}(E1)|^2 = (2J_m+1)B_{mn}(E1)$ of the E1 transitions
% electromagnetic dipole transition probabilities B$_{mn}$ 
between these excited states and the state $|m\rangle = 2_1^+$ with the peaks folded by $\Delta = 20$ keV. Panel (d) contains the 
 %isovector giant dipole resonance, the high-energy, broadly distributed dipole collective mode associated with the out-of-phase oscillation %of the neutron and proton Fermi liquids against each other.
%The 
IVGDR, which consists of a myriad of overlapping dipole transitions mainly lying in the continuum.
% and serves as a major testbed for theoretical approaches. 
It was computed in REOM$^{1-3}$ with
%the REOM hierarchy of approximations with growing correlation content and 
a typical continuum width $\Delta = 200$ keV to benchmark the theory in a broad energy region. 
REOM$^3$ showing the best performance here and in the preceding work \cite{novak2024}, was thus employed as the main tool for describing PDR$^{\ast}(2^+_1)$
%the $(1,2,3)^-$ excited states below and around the particle emission threshold and their $\gamma$ decay to the $2_1^+$ state, which is quantified 
shown in panels (a-c). 
%The theory yields several states of each of these multipolarities below 10 MeV. 
The first theoretical peak is seen in the 
%natural-parity 
$3^-$ channel at E = 4.53 MeV and identified as the lowest collective isoscalar octupole vibration,
% known in many nuclei. 
%\textcolor{blue}{
%This state likely
%usually is not associated with the PDR and 
likely corresponding to the observed excitation at 4.324 MeV with 50(1)\% $\gamma$-branching ratio to $2_1^+$, also predominantly isoscalar. Since the theory is not exact, we anticipated an energy mismatch of around a few hundred keV between theory and experiment. Therefore, we hypothesize that the observed resonant states at 7.840 and 7.996 MeV correspond to the theoretical $3^-$ excitations at 8.33 and 8.43 MeV
%, which exhibit 
%\textcolor{blue}{
with significant transition probabilities $B_{mn}(E1)$ to $2_1^+$
and similar ratios of their $B_{mn}(E1)$ values and energies. The group of $J^{\pi} = (1,2)^-$ states between 7.5 and 9 MeV show notably smaller $B_{mn}(E1)$ to the $2_1^+$ state.
%unnatural-parity $2^-$ states at 8.36 and 8.68 MeV 
%may also be candidates for being the observed 
%PDR$^{\ast}(2^+_1)$ 
%peaks, but the probabilities of their E1 transitions to the $2_1^+$ state are notably smaller.
For the 
%total reduced transition probabilities B$_{mn}$(E1) 
summed squared matrix element 
%B$_{mn}$(E1) 
$|{\bar F}_{mn}(E1)|^2$
below 10 MeV, we obtained 
%0.47$\times$3 = 
1.41 
[e$^2$ fm$^2$] for the $0^+(g.s.)\leftrightarrow 1^-$ 
%(factor 3 takes into account the degeneracy of the $1^-$ states) 
as compared to 0.013, 0.016, and 0.266 [e$^2$ fm$^2$] for the $2^+_1\leftrightarrow (1,2,3)^-$ 
%(no degeneracy included in the definition) 
transitions, respectively, i.e., 
%. Thus, for the reduced transition probabilities, before accounting for the level degeneracies, we obtain 
%comparable 
smaller but still substantial total strengths for PDR$^{\ast}(2^+_1)$ as compared to PDR. 
%Further details, including the corresponding running sums and multiplicity factors, are given in the section "Methods: Theoretical approach" and Fig. \ref{RunSum}. 
%}
%%%%%%%%%% 

Panels (e-h) of  Fig. \ref{theory} feature the transition densities in the specified energy intervals. For the $1^-$ states, the transition densities $\delta\rho^{0n}$ between the ground and excited states were extracted for both PDR and IVGDR and exhibit the behavior analogous to those exemplified in Fig. \ref{fig0}(e). The IVGDR transition densities around its centroid manifest the canonical out-of-phase oscillation of neutrons against protons and are shown in Fig. \ref{theory}(h), while the PDR ones 
%exhibit in-phase oscillations with some neutron dominance on the nuclear surface. 
%\textcolor{blue}{
representing its energy evolution are given in Fig. \ref{TrDen_all} of the section "Methods: Theoretical approach". Since in this work we focus on PDR$^{\ast}(2^+_1)$, the relevant characterization of its underlying structure is the generalized transition density connecting the two excited states $|m\rangle$ and $|n\rangle$. Such transition densities denoted by $\delta\rho^{mn}\equiv \delta\rho^{\ast}$ were determined by the angular momentum recoupling of the reduced matrix elements of $\delta\rho^{0m}$ and $\delta\rho^{0n}$, where the index "0" marks the ground state $|0\rangle$: $\delta\rho^{mn}(1^-) = \delta\rho^{nm}(1^-) = [\delta\rho^{0m}(J_m^{\pi_m})\otimes \delta\rho^{0n}(J_n^{\pi_n})]_{1^-} $ for $J_m^{\pi_m} = 2^+$ and $J_n^{\pi_n} = (1,2,3)^-$, as detailed in the section "Methods: Theoretical approach".

An immediate observation is that the behavior of the transition densities $\delta\rho^{mn}(1^-)$ below 10 MeV is firmly distinct from the IVGDR pattern, which unambiguously qualifies all the transitions as PDR$^{\ast}(2^+_1)$. A more delicate task is to establish how $\delta\rho^{mn}(1^-)$ evolves within its energy span, for which we extracted $\delta\rho^{mn}(1^-)$ for all the dipole transitions $2^+_1 \leftrightarrow (1,2,3)^-$, also collected in Fig. \ref{TrDen_all}. 
Although $\delta\rho^{mn}(1^-)$ is generally different for different channels, the common feature is the change of the relative behavior of its proton and neutron components around 9 MeV. To illustrate these structural differences, we have selected {\it characteristic} states with the maximally expressed $2q$ content within the 1 MeV intervals below and above 9 MeV.
%, which we refer to as characteristic transition densities. The transition densities $\delta\rho^{mn}(1^-)$ associated with these states are marked as "l" and "u" for the $2^+_1 \leftrightarrow (1,2,3)^-$ channels of PDR$^{\ast}(2^+_1)$ and displayed in Fig. \ref{theory} (e-g), respectively.
For the dominant $2^+_1 \to 3^-$ channel as well as for the $2^+_1 \to 2^-$ one, the lower ("l") part favors the isoscalar type of excitation, %which manifests coherent oscillations of protons and neutrons around the almost "frozen" quadrupole shape associated with the low-%frequency 2$_1^+$ state, 
while the upper ("u") part develops a well-pronounced neutron dominance in the outer space, manifesting the strong signature of a neutron-skin oscillation. 
%In REOM$^3$, the separation energy between the two components is approximately 9 MeV; however, this value may be somewhat model-%dependent. 
%the effect of the two-component splitting, 
% In the lower-energy part, all the transition densities peak around 5 fm, 
%indicating the coherent change in shape occurring during the transition. The 
%while the higher-energy interval favors the neutron-skin oscillation, which is seen as an extended neutron matter distribution beyond the %nuclear surface in the 2$^-$ and 3$^-$ cases. Further observations indicate some peculiarities, in particular, that (i) the transitions to the %2$^-$ states are single-isospin dominated because of the overall weak isospin mixing in unnatural-parity states, (ii) 
The $2^+_1 \leftrightarrow 1^-$ transition densities for the $1^-$states around 9 MeV rather exhibit some decoherence and increasing proton contribution because of the IVGDR admixture.
%, and (iii) in all the channels $J^{\pi} = (1,2,3)^-$, the generalized $2^+_1 \to J^-$ transition densities retain similarities with the regular %g.s.$ \to J^-$ transition densities because the latter ones are obtained by the recoupling of the former ones with the isoscalar g.s.$ \to %2^+_1$ transition densities.
%%%%%%% 
Thereby, PDR$^{\ast}(2^+_1)$ essentially retains the underlying features but also exhibits some structural differences from the ground-state-based PDR. 
%Particularly, while those to the 3$^-$ states inherit the isoscalar and neutron-skin patterns, which are present in the ground-state-based PDR %anatomy elaborated in the insets of Fig. \ref{fig0}(e). These features are consistent with the definition of the recoupled transition density 
%$\delta\rho^{mn}(1^-)$ as a tensor product of the two regular transition densities: one between the ground and the $2^+_1$ state and another %between the ground and the $(2,3)^-$ states. 
%%% here may be a remark about higher energy.
%This composition is remarkably consistent with the two-component structure of the ground-state-based PDR \cite{endres2010isospin,Markova2024} and %its distinction from the GDR.

%\section*{Online content}

%Any methods, additional references, Nature Portfolio reporting summaries, source data, extended data, supplementary information, acknowledgements, peer review information; details of author %contributions and competing interests; and statements of data and code availability are available.

%\bibliography{80Ge-bibliography.bib}

\section*{Methods}
\label{method}

\subsection*{Precursor identification}
\label{precursor_identification}
For obtaining two separated decay level schemes of $^{80g}$Ga and $^{80m}$Ga, it is essential to identify the $\beta$-feeding precursor for each state of the daughter nucleus. The method we use is based on the measurement of each individual $\gamma$-line apparent half-life, i.e., the so-called "X" value method \cite{verney2013structure,Li2025a}. $\beta$-decay of $^{80m}$Ga with $J^{\pi} = 3^-$ would primarily populate states of spin-parity (2,3,4)$^-$ through allowed transitions and those of (2,3,4)$^+$ through first-forbidden non-unique (ffnu) transitions. Similarly, the (5,6,7)$^-$ and (5,6,7)$^+$ states are primarily populated by $^{80g}$Ga with $J^{\pi} = 6^-$. Therefore, there is no overlap between the states directly $\beta$-fed by the two isomers if only the allowed and ffnu transitions are considered. An overlap would be only possible via first-forbidden unique (ffu) transitions with logft > 8, according to the systematic analysis in the A = 80 region \cite{singh1998review}. So, if ffu occurs, the crossed $\beta$-feeding can happen to the 5$^+$ and 4$^+$ states.
%}

%
\begin{figure*}
    \centering
    \includegraphics[width=0.7\textwidth]{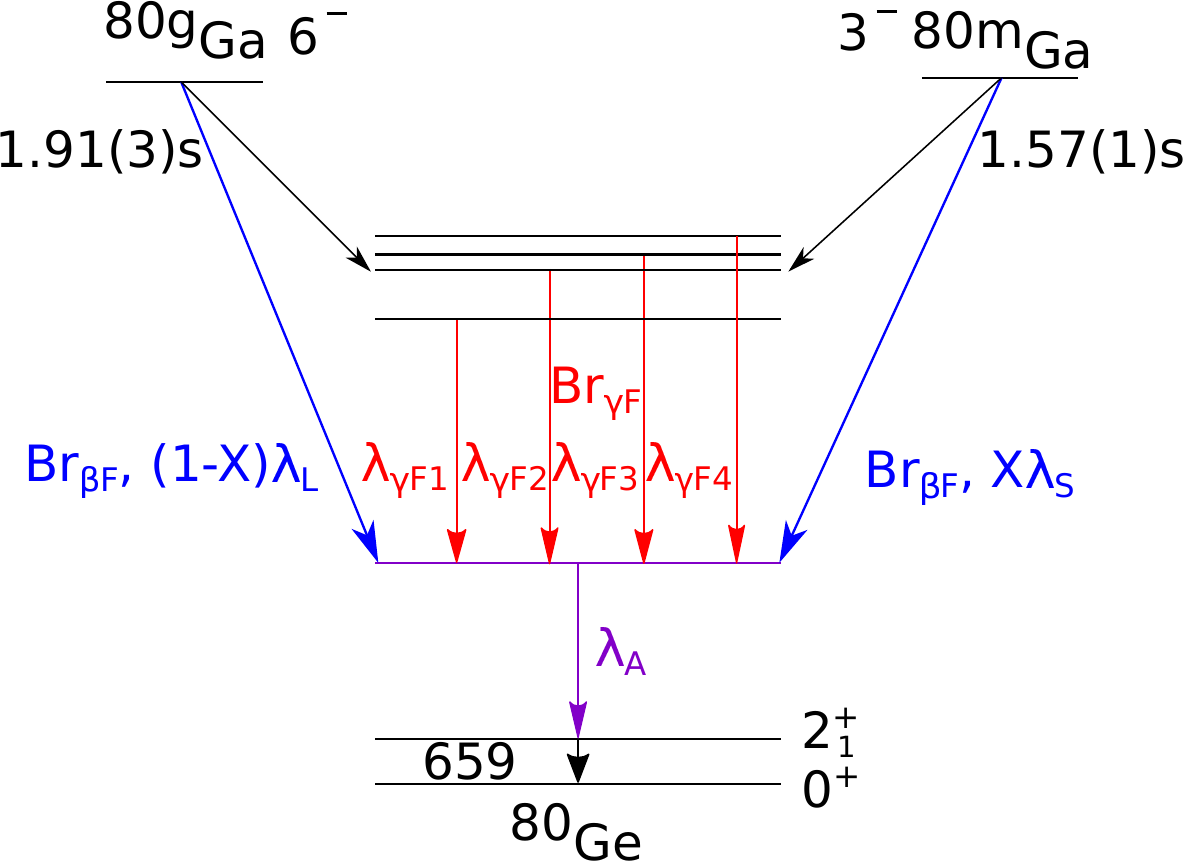}
    \caption{Schematic diagram of the "X" equation: the parameters involved in building the "X" equation for a given state, which is marked in purple.}
    \label{fig6.20}
\end{figure*}

\begin{figure*}%[!htb]
\centering
\includegraphics[width=0.80\textwidth]{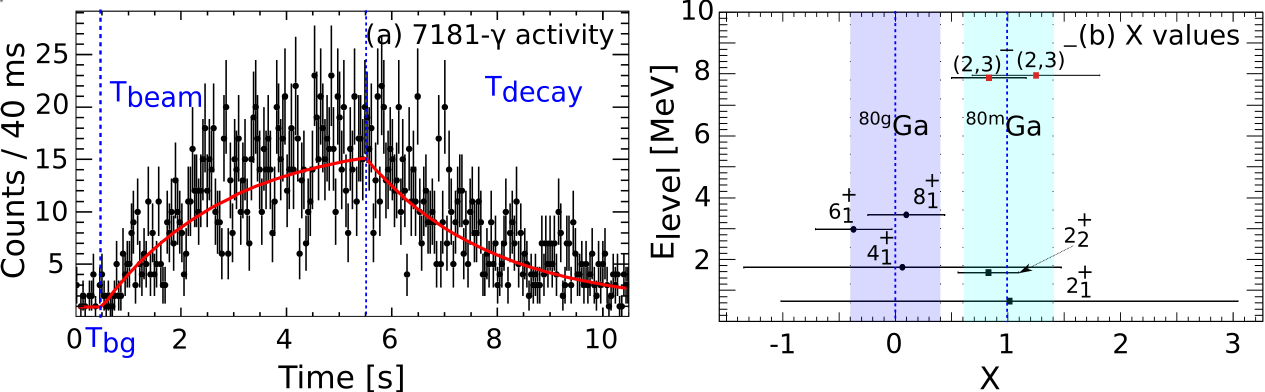}
\caption{(a): 7181 keV $\gamma$-ray activity curve and its fit; (b): X values for the 7840 keV and 7996 keV states together with the values for well-identified states,
%. X is fractional direct $\beta$-feeding from $^{80m}$Ga (Br$_{{\beta}F}$), see more 
see details in the Section 
%\ref{method} 
"Precursor identification".}
\label{fig2}
\end{figure*}

A quantity "X" is proposed to help in assessing the belonging of the excited states of $^{80}$Ge to one of the $^{80g}$Ga and $^{80m}$Ga decay schemes. For each excited state, one determines an apparent half-life (T$_{1/2}^A$) from a specific $\gamma$-ray activity curve like the one shown in Fig. $\ref{fig2}$(a) or a weighted value of a few of $\gamma$-rays emitted from the same state in $^{80}$Ge. The value of this apparent half-life is the result of several contributions: the fractional direct $\beta$-feeding from $^{80g}$Ga (Br$_{{\beta}F}$(1-X)), fractional direct $\beta$-feeding from $^{80m}$Ga (Br$_{{\beta}F}$X), the half-life of $^{80m}$Ga (T$_{1/2}^S$), the half-life of $^{80g}$Ga (T$_{1/2}^L$), the branching ratio of $\gamma$-feeding (Br$_{{\gamma}F}$), and the weighted half-life of $\gamma$-rays taking part in this $\gamma$-feeding (${\overline T}_{1/2}^{{\gamma}F}$). Fig. $\ref{fig6.20}$ presents the schematic diagram of the relationships between these parameters, which is expressed by the following equation:

\begin{equation} \label{X1} 	
\lambda_A = \frac{Br_{{\beta}F}}{Br_{{\beta}F} + Br_{{\gamma}F}}[X\lambda_S + (1-X)\lambda_L] + \frac{Br_{{\gamma}F}}{Br_{{\beta}F} + Br_{{\gamma}F}}\lambda_{{\gamma}F},
\end{equation}

where $\lambda_{A/S/L}$ = $\ln 2/T_{1/2}^{A/S/L}$, T$_{1/2}^A$ is the apparent half-life of a given excited state and ${\lambda}_{{\gamma}F} = \ln 2/{\overline T}_{1/2}^{{\gamma}F}$ is the 
%\sout{apparent} 
decay constant associated with the $\gamma$-feeding, 
the so-called observed indirect feeding. T$_{1/2}^S$ and T$_{1/2}^L$ are the new measured values for $^{80m}$Ga and $^{80g}$Ga, which are 1.57(1) s and 1.91(3) s, respectively, and Br$_{{\beta}F}$ + Br$_{{\gamma}F}$ = 1 for a given state in $^{80}$Ge. The equation for X reads:

\begin{equation} \label{X2} 	
X = \frac{1}{R}\frac{1/{\overline T}_{1/2}^{{\gamma}F} - 1/T_{1/2}^A}{1/T_{1/2}^L - 1/T_{1/2}^S} + \frac{1/T_{1/2}^L - 1/{\overline T}_{1/2}^{{\gamma}F}}{1/T_{1/2}^L - 1/T_{1/2}^S}, 
\end{equation}

where $\overline{T}_{1/2}^{{\gamma}F}$ was taken as the weighted average of all the $\gamma$-rays
% of indirect feeding as shown in Eq. (\ref{X3}) 
with the associated error: 
%\textcolor{red}{in} 
%Eq. (\ref{X4}) below.
%

\begin{equation} \label{X3} 	
	\overline{T}_{1/2}^{{\gamma}F} = \sum_{i=1}^n Br_{{\gamma}Fi}T_{1/2({\gamma}i)}, \ \ \ \ \ \ \ \ \ \ \ 
%\end{equation}
%
%\begin{equation} \label{X4} 	
	\overline{{\Delta}T}_{1/2}^{{\gamma}F} = [\sum_{i=1}^n ({\Delta}Br_{{\gamma}Fi}T_{1/2({\gamma}i)})^2 + (Br_{{\gamma}Fi}{\Delta}T_{1/2({\gamma}i)})^2]^{\frac{1}{2}}.
\end{equation}

$R$ is the proportion 
 of the direct $\beta$-feeding contribution in the total direct and indirect feeding of one given state, calculated directly from the difference between the $\gamma$-ray counts of de-excitation and $\gamma$-feedings as follows: 
 %Eq. $\ref{X5}$. 
%

\begin{equation} \label{X5} 	
	R = \frac{C_{{\gamma}-from-{\beta}-feeding}}{C_{all-{\gamma}-of-deexcitation}}.
\end{equation}
It is important to point out that the 
%precision 
accuracy of the extracted X values relies primarily on the available statistics, i.e., on the capability to detect all indirect feeding strength, including that at higher energy.
% \textcolor{red}{region}.
Then, the levels with the values X = 0 are considered as perfect candidates to the decay scheme of the longer-lived $^{80g}$Ga, and X = 1 to the decay scheme of the shorter-lived isomer $^{80m}$Ga. Besides the unambiguous situations, the 
%actual actions taken are: 
actual identifications were: 1) X $<$ 0.4 belong to the ground state decay, 2) X $>$ 0.6 belong to isomer decay, and 3) 0.4 $<$ X $<$ 0.6 belong to both. 
%This standard was taken for the purpose of achieving as precisely as possible the isomer ratio, precursor absolute decayed population of %$^{80g}$Ga and $^{80m}$Ga, and the related logft and B(GT) values. 
After this analysis, only twelve states were assigned to simultaneous direct $\beta$-feeding from both $^{80g}$Ga and $^{80m}$Ga, out of a total of 81 populated states. Additionally, there are two complementary techniques for precursor assignment \cite{ren2022thesis}. As shown in Fig. $\ref{fig2}$(b), this method was validated with states of well-established spins-parities. Fig. $\ref{fig2}$(a) presents the activity curve of the 7181 keV $\gamma$-line and its fit with the solution 
%function 
of the Bateman equations.

%For the assignment of the multipolarity of $\gamma$ transitions and spin-parity to the levels, one needs to disentangle the two $^{80g}$Ga and $^{80m}%$Ga $\beta$-decay schemes. This is achieved by using a method based on the measurement of the half-lives of $\beta$-delayed $\gamma$-rays. The %apparent half-lives have two contributions: 1) 
%ratio of the 
%direct $\beta$-feedings from two isomers characterized by their intrinsic $\beta$ half-lives; 2) indirect $\gamma$-feeding from the depopulation of higher-%lying states with their own apparent half-lives. To each level, we attributed a quantity $X$, characterizing the fraction of the direct feeding from the 3$^-$ %isomer, that includes these 
%different 
%contributions and that should take a value 0 (1) if the level is populated by the $\beta$-decay of $^{80g}$Ga ($^{80m}$Ga) %\cite{verney2013structure,ren2022thesis}. 
%\textcolor{red}{
The extracted half-lives of the $\beta$-delayed $\gamma$-rays of 7181 keV and 7337 keV are 1.62(10) s and 1.50(14) s, respectively, and the derived $X$ values for the 7.84 MeV and 7.996 MeV states are 0.83(33) and 1.24(56), respectively. 
%Note that the simultaneous direct feeding from both isomers can occur only if one of the two $\beta$ transitions is of the first-forbidden unique (1U) nature. 
Because of the lower probability of ffu compared to allowed and ffnu transitions \cite{singh1998review}, the condition taken for cross $\beta$-feeding is that the $X$ value is in the range of 0.4 - 0.6. Furthermore, for allowed transitions, the candidate high-lying states populated by the $J^{\pi} = 6^{-}$ $^{80g}$Ga are (5,6,7)$^{-}$ whose probabilities of de-excitation by $\gamma$ emission to a $J^{\pi} = 2^{+}$ state are too small. Therefore, the 7.84 MeV and 7.996 MeV states are assigned to the $J^{\pi} = 3^{-}$ $^{80m}$Ga $\beta$ feeding confidently. 
%In conclusion, because of the large $\Delta J = 3$ spin difference between the two $^{80}$Ga $\beta$-decaying states, the established half-lives and $X$ values constitute reliable evidence for a direct population of the resonant states at 7.84 MeV and 7.996 MeV by the 3$^-$ isomer exclusively.

%

% \textcolor{red}{
For the extraction of the absolute $\beta$-branching ratios $I_\beta$, the $\beta$ precursor abundance was obtained directly by analyzing the total $\beta$ particle activity, relying on the fact that the Ga beams did not suffer from any contaminant activity. By solving the Bateman equations involving the activities of the five nuclei of the $^{80g+m}$Ga decay chain and normalizing to the recorded $\beta$ activity, we obtained the beam intensity and then the quantities $I_\beta$, logft, and B(GT). Once the separated decay level schemes were constructed, the 3$^{-}$ and 6$^{-}$ $\beta$-feeding statistics
%. Based on this, 
and the beam ratio of the two isomers were obtained, resulting in 37(2)$\%$ of the 3$^-$ isomer in the beam. The extracted absolute $I_\beta$ values for the 7.84 MeV and 7.996 MeV states are 1.01(11)$\%$ and 0.31(3)$\%$, and the related logft are 4.90(6) and 5.29(6), respectively. For the first-forbidden $\beta$ transitions to the well-established levels in $^{80}$Ge observed in the present work, logft values are all larger than 6.3. This is consistent with Ref. \cite{singh1998review} where 98$\%$ of the first-forbidden $\beta$ transitions have the logft values larger than 6. Based on this analysis, the spin-parity characteristics of the two resonant excited states can be narrowed to $(2,3,4)^{-}$.
%}

\subsection*{Multiplicity of high-energy $\gamma$-rays and background assessment}

\begin{figure}
    \centering
    \includegraphics[width=0.8\columnwidth]{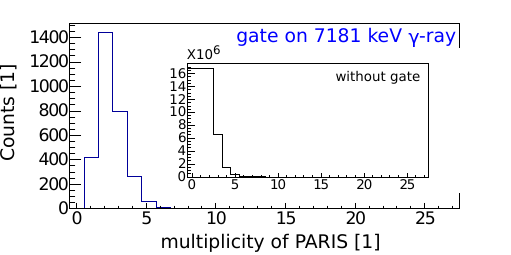}
    \caption{Multiplicities registered by PARIS 
    %\textcolor{red}{with} 
    gating on the 7181 keV $\gamma$-ray and without gating.}
    \label{fig6.21}
\end{figure}

Fig. $\ref{fig6.21}$ presents the multiplicity registered by PARIS gating on the 7181 keV $\gamma$-ray. The maximum value is 2. Note that it is different from the multiplicity of PARIS without any gate wherein the maxima are 0, 1, and 2, as shown in the inset of Fig. $\ref{fig6.21}$. It proves that the cascade containing the 7181 keV $\gamma$-ray has two $\gamma$-lines. This is in agreement with Figs. \ref{fig1}(b) and (c) of the article showing the coincidence between the 7181 keV and 659 keV $\gamma$-lines. The maximum value of the multiplicity of PARIS gating on the 7337 keV $\gamma$-ray is 2 as well. Note that Fig. \ref{fig6.21} is not constrained by detecting one 659 keV $\gamma$-line in HPGe, in contrast to Fig. \ref{fig1}(b). Therefore, the multiplicity of PARIS gated on the 7181 keV or 7337 keV $\gamma$-ray is 2, as PARIS recorded more 659 keV $\gamma$ events than the HPGe detectors. 

Note that the statistics below the two peaks in Fig. \ref{fig1}(b) do not represent the background levels at 7181 and 7337 keV, as these two peaks are located at the endpoint of the spectrum. The former is contributed by single- and double-escape and Compton-escape photons of 7181 and 7337 keV $\gamma$-rays, while the latter by $\gamma$-ray(s) above 7337 keV. One finds that the statistics above the 7337 keV peak is very low, i.e., 2(1) events per energy range of 200 keV. Consequently, the background in the 7181 and 7337 keV peaks caused by higher-energy $\gamma$-ray(s) is very low. Finally, the Compton edge of 7337 keV $\gamma$-ray is 7091 keV, just 
%before
below the edge of the 7181 keV peak. Therefore, there is no Compton-caused background from the 7337 keV $\gamma$-ray in the 7181 keV peak. The energy uncertainties include (i) the statistical uncertainty that depends on the 
%statistics 
cumulative yield of the $\gamma$-peak and is dominated by the energy resolution of the detector, and (ii) the systematic error 
%that is included in the errors of 
originating from the energy calibration of the detector.
%} 

\subsection*{Pulse-shape discrimination of PARIS}

PARIS \cite{PARIS} was designed as a calorimeter for measurements of $\gamma$-rays covering a wide energy range from 100 keV - 50 MeV. The PARIS array consists of phoswich detectors, 
each of which has two shells of scintillator detectors. 
%can provide 
The inner shell of LaBr$_3$(Ce) crystals offers an optimal performance in providing fast timing, direction, $\gamma$-ray multiplicity, and total absorption spectrum (TAS) measurement. The total energy of a high-energy photon, if it punches through and forms an electromagnetic shower in the outside shell, can be reconstructed in an add-back mode, exploiting the energy information from the outer shell, or can be screened by anti-coincidence operation. Therefore, it has the ability to distinguish full energy deposit peaks in LaBr$_3$(Ce) and NaI(Tl), cross-talking peaks, and the most important pile-up events, which is difficult in traditional TAS.

\begin{figure}
    \centering
    \includegraphics[width=0.8\columnwidth]{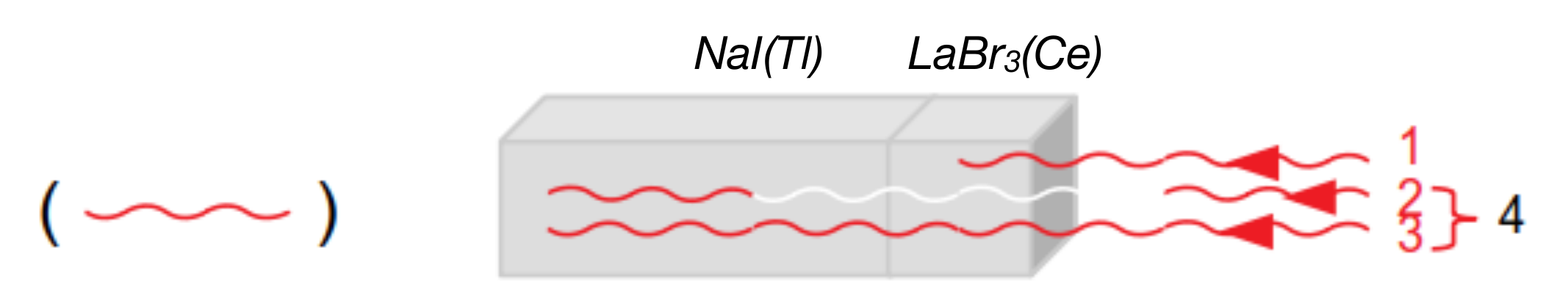}
    \caption{Energy deposit situations when a $\gamma$-ray hits one phoswich.}
    \label{fig6.24}
\end{figure}

\begin{figure*}
\centering
\includegraphics[width=0.7\columnwidth]{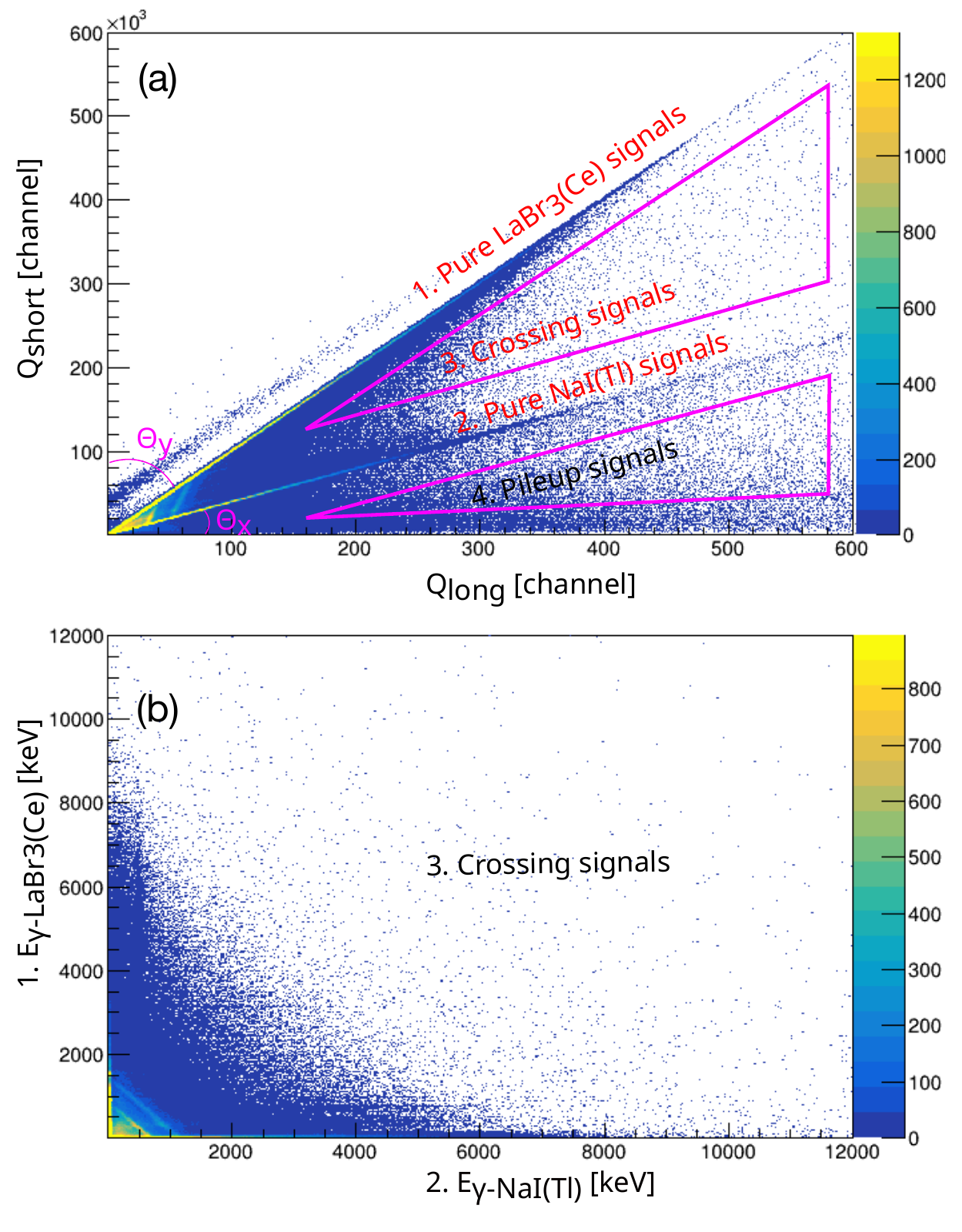}
\caption{
%\textcolor{blue}
%{
2D histogram of the two output signals $Q_{short}$ vs $Q_{long}$ from one phoswich (a) and its projection histogram of $E_{LaBr_3(Ce)}\equiv E_{\gamma-LaBr_3(Ce)}$ vs $E_{NaI(Tl)}\equiv E_{\gamma-NaI(Tl)}$ (b). Region 1: pure LaBr$_3$(Ce) signals which present the $\gamma$-rays events depositing full energy in a LaBr$_3$(Ce) crystal (after rejecting the non-full energy peaks and pileup events); Region 2: pure NaI(Tl) signals of the $\gamma$-ray events depositing full energy in a NaI(Tl) crystal; Region 3: crossing signals from the $\gamma$-ray events depositing energy partly in both crystals; Region 4: random pileup signals of more than one $\gamma$-ray hitting one phoswich simultaneously (120-820 ns time window).
%, random pileup displayed in Fig. \ref{fig6.24}. 
%It is worth noting that region 1 presents events after rejecting the non-full energy peaks and rejecting pileup events, as shown in Fig. \ref{fig1}(a) %and Fig. \ref{fig1}(b).
%}
}
\label{fig6.25}
\end{figure*}

The first step of using PARIS is to separate the signals coming from LaBr$_3$(Ce) and NaI(Tl), which is essential for their individual calibrations. Pulse-shape discrimination 
%(PSD) 
methods are used to resolve the two phoswich signals using their widely different shaping and decay times, as shown in Fig. 2 of Ref. \cite{ghosh2016characterization}. So, one can save two values generated by the FASTER DAQ \cite{FASTER} system: one was obtained with a shorter time gate of 120 ns for the charge-to-digital conversion (QDC) process, and the second one 
%was obtained 
with a longer time gate of 820 ns for the QDC coding window.
%
%\begin{figure}
%    \centering
%    \includegraphics[width=0.6\columnwidth]{FIG8.pdf}
%    \caption{The pulses of signals in one phoswich generated by a high-energy $\gamma$-ray in a test measurement of PARIS. The figure is adapted from Ref. \cite{ghosh2016characterization}.}
%    \label{fig6.23}
%\end{figure}
%
There are 
%more than 
the following different situations when a $\gamma$-ray arrives in the phoswich as illustrated in Fig. $\ref{fig6.24}$:
\begin{enumerate}
\item 
The $\gamma$-ray hits the LaBr$_3$(Ce) crystal only and deposits its full energy inside. In this case, one can observe the same value of charge collected in a photomultiplier tube Q$_{short}$  with a shorter QDC time gate and Q$_{long}$ with a longer time gate.
%, as shown in Fig. $\ref{fig6.24}$. 
This situation is characterized by the angle $\theta_y$ 
%= 45^{\circ}$ 
in Fig. \ref{fig6.25}(a); 
\item 
The $\gamma$-ray goes through LaBr$_3$(Ce) but deposits zero energy and then hits NaI(Tl) and deposits full energy there. This gives rise to another clear line in the diagram of Fig. \ref{fig6.25}(a) characterized by the angle $\theta_x$; 
\item 
The $\gamma$-ray punches through LaBr$_3$(Ce) and NaI(Tl) and deposits energy in both crystals partly. These events are localized in the crossing part of Fig. \ref{fig6.25}(a); 
\item
Two or more $\gamma$-rays hit the phoswich at the same time. These are pile-up events, which appear only in the longer QDC time window of 820 ns. In that situation, Q$_{long}$ is larger than usual, as seen in region 4 of Fig. \ref{fig6.25}(a). Note that these data are a pile-up from random-time events, while the $\gamma$-cascade events do not appear in this region. However, this effect is very small due to the large granularity of PARIS and low multiplicity in the $\beta$-decay level scheme; 
\item
The photon escapes the phoswich from the bottom or side direction. 
%This situation is more complex. 
One can remove part of the bottom escape events from the spectra by using NaI(Tl) as veto detectors. In the case of the side escape, one can get rid of these events by a cluster add-back procedure.
\end{enumerate}

On the next step, the measured values $Q_{short}$ and $Q_{long}$ were used to generate the quantities $q_1(E_{LaBr3(Ce)}$) and $q_2(E_{NaI(Tl)}$), which contain the LaBr$_3$(Ce) and NaI(Tl) energy deposit information exclusively:

%\textcolor{green}{
\begin{align} \label{PARIS-projection} 	
	&\begin{aligned}
	& Q_{short} & = q_1(E_{LaBr_3(Ce)})\cos{\theta}_y + q_2(E_{NaI(Tl)})\sin{\theta}_x	\\ 
	& Q_{long}  & = q_1(E_{LaBr_3(Ce)})\sin{\theta}_y + q_2(E_{NaI(Tl)})\cos{\theta}_x, 		
	\end{aligned}	
\end{align}
%
%\textcolor{blue}{
employing the two parameters $\theta_x$ and $\theta_y$. Then, $E_{LaBr_3(Ce)}$ and $E_{NaI(Tl)}$ can be obtained by calibrating $q_1(E_{LaBr_3(Ce)}$) and $q_2(E_{NaI(Tl)}$) using the reference $\gamma$ sources. Fig. \ref{fig6.25}(b) shows the result of this procedure. 
%At the moment, the $\gamma$ spectra from LaBr$_3$(Ce) and NaI crystals are obtained.
%}
The pure LaBr$_3$(Ce) (NaI(Tl)) $\gamma$ spectrum is then obtained by setting $E_{NaI(Tl)}$ = 0 ($E_{LaBr_3(Ce)}$ = 0) and the LaBr$_3$(Ce)-NaI(Tl) add-back spectrum by adding the LaBr$_3$(Ce) and NaI(Tl) signals. Using this add-back procedure allows recovering the full energy peaks part of the statistics spread over the 511 keV single- and double-escape events. This is particularly important for the highest energy part of the $\gamma$ spectrum. Furthermore, PARIS has the ability to distinguish neutrons and $\gamma$ through their time of flight difference at a distance of 15 cm.

\subsection*{The $\gamma$ decay pattern of the $J^{\pi} = (2,3)^-$ high-energy components of PDR$^{\ast}$(2$_1^+$) }

We also observed a few $\gamma$ transitions 
%from the high-lying PDR$^{\ast}$(4$_1^+$) 
%$J^{\pi} = (2,3)^-$ 
%candidates 
from the high-lying PDR$^{\ast}$ candidates at 7.84 MeV and 7.996 MeV 
to the 4$_1^+$ state \cite{ren2022thesis}. However, no decay of the 7.84 MeV and 7.996 MeV states directly to the second quadrupole 2$_2^+$ state has been registered. From a macroscopic point of view, $^{80g}$Ga and $^{80m}$Ga have the maximum electric quadrupole moments among the $^{65,67,69,71,75,79-82}$Ga isotopes \cite{farooq2017probing}. Consequently, the 7.84 MeV and 7.996 MeV states must also have substantial quadrupole components, consistent with the precursor $^{80m}$Ga. The significant overlap between these states and 2$_1^+$ results in the preference to decay to 2$_1^+$ with a larger quadrupole deformation $\beta_2$ = 0.155(9) than that of the 2$_2^+$ state ($\beta_2$ = 0.053$^{0.008}_{0.009}$) \cite{iwasaki2008persistence}. From a microscopic perspective, the larger overlap of the wave functions of the 7.84 MeV and 7.996 MeV states and the 2$_1^+$ one, with larger amplitudes of neutron configurations \cite{padilla2005b,iwasaki2008persistence}, can be used to explain the preferred transition between them rather than de-excitation to 2$_2^+$. In the shell model, the main component of these high-lying excited states is $(\pi1f_{5/2})^2$ $(\nu1g_{9/2})^{-1}$ $(\nu1f_{7/2})^{-1}$, i.e., a 2p2h configuration. Based on the B(E2) measurements and shell model calculations \cite{iwasaki2008persistence}, the main proton contributions to the 2$_1^+$ state are  $(\pi1f_{5/2})^2(\pi2p_{3/2})^2$ and $(\pi1f_{5/2})^4$ configurations (50$\%$ in total), whereas the 2$_2^+$ state is dominated by the $(\pi1f_{5/2})^2(\pi2p_{3/2})^2$ and $(\pi1f_{5/2})^3(\pi2p_{3/2})^1$ components, differing from the 2$_1^+$ one by a particle-hole pair.
The prevailing neutron configuration in both 2$_1^+$ and 2$_2^+$ states is $(\nu1g_{9/2})^{-2}$ (about 80$\%$), however,  the two holes tend to be coupled to 2$^+$ in the 2$_1^+$ state and to 0$^+$ in the 2$_2^+$ state. Consequently, the 7.84 MeV and 7.996 MeV states have higher probabilities of de-exciting to 2$_1^+$ via E1 transitions than to 2$_2^+$. 
%This analysis indicates that the decay pattern of PDR depends on the nuclear structure of the involved states. 
%Since two states with sizable strength could be investigated in detail in the present work due to statistical limitations, clearly, more experimental data are %required for further investigating the evolution of PDR properties as a function of excitation energy, deformation, and spin.

Preceding studies of the ground-state-based $J^{\pi} = 1^-$ PDR indicate that in medium-mass neutron-rich nuclei this mode can have fragments both below and above $S_n$  \cite{wieland2009search,rossi2013measurement,Wieland2018}. This could potentially be the case for the $J^{\pi} = (1,2)^-$ components of PDR$^{\ast}(2_1^+)$ in $^{80}$Ge, while the theory predicts only a negligible fraction of its $J^{\pi} = 3^-$ component above $S_n$. Theoretical transition densities of the $J^{\pi} = 1^-$ states above 10 MeV evolve toward the IVGDR pattern, suggesting that indeed the main part of the ground-state-based $J^{\pi} = 1^-$ PDR is below this energy. On the experimental side, excited states above S$_n$ can still be populated by the beta decay up to the Q$_{\beta}$ = 10.312(4) MeV value, where $\gamma$-neutron competition should be taken into account. 

\subsection*{Theoretical approach}
\label{methods:theoretical approach}

The REOM approach has emerged recently as a consolidation of the long-term effort on the nuclear many-body theory and its numerical implementations  \cite{PLANT1998607,PhysRevC.78.014312,PhysRevC.88.054301,PhysRevC.100.064320,PhysRevC.106.064316}. It is formulated in the universal language of quantum field theory, adopted to atomic nuclei as systems of nucleons and mesons whose quark structure is considered irrelevant to low-energy nuclear physics. The response of a strongly coupled fermionic system to external perturbation is quantified by the REOM for the in-medium propagator for a particle-hole ($ph$) or quasiparticle ($2q$) pair, obtained via the canonical Bogoliubov transformation and excited out of the ground state.  The dynamical kernel of this REOM characterizes the involvement of multiple quasiparticle pairs in the formation and decay of the excited state. Three levels of configuration complexity (i) $2q$ ($\sim 1p1h$), (ii) $2q\otimes phonon$ ($4q\sim 2p2h$), and (iii) $2q\otimes 2phonon$ ($6q\sim 3p3h$), with phonon as an emergent correlated  $2q$ pair, can be employed and compared to assess the quality of the many-body approach \cite{novak2024}. REOM$^1$ is equivalent to the relativistic quasiparticle random phase approximation (RQRPA) \cite{Paar2003} and generates "primordial" correlated $2q$ configurations, which are used as the basis for higher-complexity states obtained via the well-converging iterative procedure \cite{PhysRevC.100.064320,novak2024}. The calculations were performed with one of the latest non-linear meson-exchange interactions, NL3$^\ast$ \cite{LALAZISSIS200936}, whose parameters are universal across the nuclear chart. The pairing strength parameters, which are not part of the NL3$^\ast$, were adjusted to the experimental position of the $2_1^+$ state in $^{80}$Ge, and its measured B(E2) value was used to gauge the theoretical transition probability of this state. In the absence of experimental data on IVGDR in $^{80}$Ge, the theoretical result in Fig. \ref{theory}(d) can be assessed by the empirical position of its centroid at $E_0 = 80/\sqrt{A} \approx 18.57$ MeV, where $A = 80$  is the mass number, and the recommended width of the main peak of 6.4 MeV \cite{RIPL3}.  

%%%% moved
%\textcolor{red}{Since the REOM solutions saturate quite fast, rather moderate complexity can provide a fairly accurate description of nuclear spectra in %large energy intervals \cite{novak2024}. In open-shell superfluid nuclei, the basis of $ph$ configurations is replaced by the basis of two-quasiparticle %($2q$) pairs via the canonical Bogoliubov transformation. 
% In this work, we employ the maximally feasible REOM$^3$ version of the theory and generalize it to the transitions between excited states. 
% The $6q$ configurations organized as $2q\otimes 2phonon$, where the $phonon$ stands for a correlated $2q$ pair, is currently the maximally feasible %complexity. In the absence of experimental data on IVGDR in $^{80}$Ge, the result can be assessed by the empirical position of the centroid at $E_0 = %80/\sqrt{A} \approx 18.57$ MeV, where $A = 80$  is the mass number.  }
 %\ref{methods:theoretical approach}.
%%%%

%In this work, 
%\textcolor{red}{
We apply the approach of maximally possible $6q$ complexity dubbed as REOM$^3$, generalized to transitions between excited states, to compute the transitions $2_1^+ \leftrightarrow (2,3)^-$  observed in the reported experiment and $2_1^+ \leftrightarrow 1^-$ transitions to complete the PDR$^{\ast}(2_1^+)$ multiplet. In the leading approximation, the electric dipole transition matrix element between the two states $|m\rangle$ and $|n\rangle$ is expressed via contraction of their superfluid two-component transition densities $\delta\rho^{0k} = \{{\cal X}^{k},  {\cal Y}^{k}\}, k = \{m, n\}$, with the E1 operator $F^{(E1)}$: $F_{mn} = \langle m|F^{E1}|n\rangle = F^{E1(11)}_{\alpha\beta}({\cal X}^{m\ast}_{\alpha\gamma}{\cal X}^{n}_{\beta\gamma} + {\cal Y}^{m\ast}_{\beta\gamma}{\cal Y}^{n}_{\alpha\gamma})$ \cite{PhysRevC.106.064316,BERTULANI1999139} over the repeated Greek indices denoting Bogoliubov's quasiparticle states generated by the operators 
$\{\alpha^{\dagger}_{\beta},\alpha_{\beta}\}$, such as 
${\cal X}^{k}_{\beta\gamma} = \langle 0|\alpha_{\gamma}\alpha_{\beta}|k\rangle$ and ${\cal Y}^{k}_{\beta\gamma} = \langle 0|\alpha^{\dagger}_{\beta}\alpha^{\dagger}_{\gamma}|k\rangle$. The upper index $(11)$ in the dipole operator marks its Bogoliubov's component \cite{RS80}.
The observable of interest is the squared reduced matrix element %$B_{mn}(E1) = |{\bar F}_{mn}|^2$ expressed via the reduced transition 
matrix element \cite{Edmonds} ${\bar F}_{mn} = \langle m||F^{E1}||n\rangle = F^{E1(11)}_{(\alpha\beta)}\delta\rho^{mn}_{(\alpha\beta)}(1^-)$, independent of the direction of the transition: ${\bar F}_{mn} = {\bar F}_{nm}$.
%Coupling $[2_1^+\otimes 1^-]$ generates a $(1,2,3)^-$ multiplet, whose major $(2,3)^-$ low-energy part (below $\approx$ 10 MeV) is identified as the %observed PDR$^{\ast}(2_1^+)$. 
%
%
%Within the self-consistent REOM$^3$ approach, 
%
%\textcolor{red}{
The $(1,2,3)^-$ states under study 
consist mainly of $4q$ and $6q$ configurations, fragmented from the primordial $2q$ REOM$^1$ modes, and the $2_1^+$ state 
is dominated by the one-phonon ($2q$) contribution. In the $2^-$ states,
the static residual interaction mediated by unnatural-parity mesons is weak
 \cite{PhysRevC.36.380}, explaining weak mixing of proton and neutron contributions in such states within the present approach, where the static unnatural-parity residual interaction was neglected. 
\begin{figure}
    \centering
    \includegraphics[width=0.9\columnwidth]{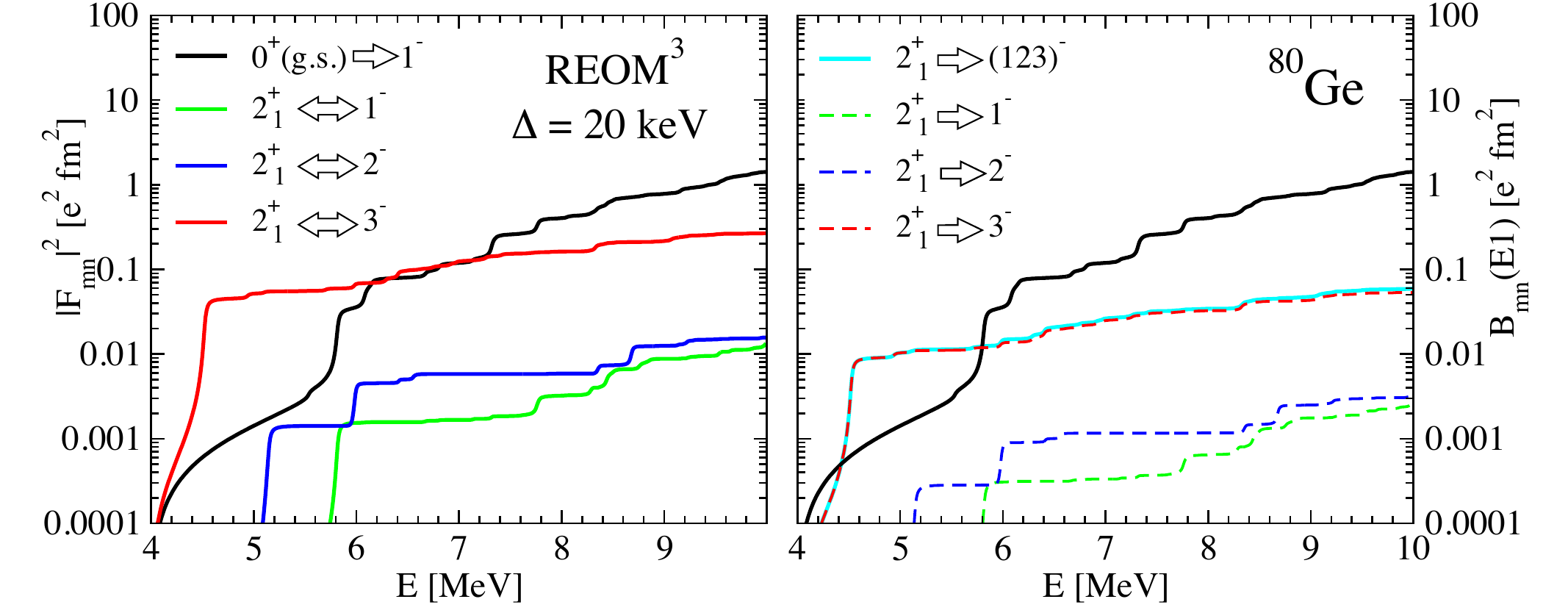}
    \caption{Running sums for the squared reduced matrix elements $|{\bar F}_{mn}(E1)|^2$ (left) and $\gamma$-absorption reduced transition probabilities $B_{mn}(E1\uparrow)$ (right) of PDR and PDR$^{\ast}(2_1^+)$. The light blue line shows $B_{mn}(E1\uparrow)$ for PDR$^{\ast}(2_1^+)$ summed over the three spins.}
    \label{RunSum}
\end{figure}

%\textcolor{red}{
Fig. \ref{RunSum} shows the running sums for $|{\bar F}_{mn}(E1)|^2$ and absorption $B_{mn}(E1\uparrow)$ of the ground-state-based PDR and the $J^{\pi} = (1,2,3)^-$ components of the PDR$^{\ast}(2_1^+)$ obtained in REOM$^3$. The $\gamma$ emission and absorption probabilities are defined as $B_{mn}(E1) = |{\bar F}_{mn}(E1)|^2/(2J_m+1)$, where $J_m$ is the spin of the initial state \cite{bohr1998nuclear}. 
Note that for the ground-state-based PDR $B_{mn}(E1\uparrow) = |{\bar F}_{mn}(E1)|^2$, while the absorption $B_{mn}(E1\uparrow)$ of PDR$^{\ast}(2_1^+)$ is suppressed by the factor of five, the degeneracy of the $2_1^+$ state.  The dominance of the octupole component is evident and can be attributed to the enhanced collectivity of the $J^{\pi} = 3^-$ states. While the $J^{\pi} = (1,2)^-$ states below 10 MeV represent minor fractions of the respective total strength distributions below the giant resonances, the octupole strength in this energy range is substantially larger than in its high-energy part, in agreement with Ref. \cite{Ma2002}.
%Furthermore, the obtained $B_{mn}(E1)$ for the $2_1^+\to 3-$ transitions of the order of $10^{-2}$ [e$^2$fm$^2$] 
The $B_{mn}(E1)$ values of the resonant $2_1^+\to 3^-$ transitions of the order of $10^{-3}-10^{-2}$ [e$^2$fm$^2$] suggest that the $3^-$ component of PDR$^{\ast}(2_1^+)$ may be a dynamical analog of the phenomenon of enhanced octupole correlations in nuclei expressed via enhanced E1 transitions between excited states \cite{Alhassid1982,Rugari1993,Butler2019}. 
%}

%\textcolor{red}{
Comparison between the $B_{mn}(E1\downarrow)$ values of PDR and PDR$^{\ast}(2_1^+)$ for $\gamma$ emission is less unambiguous as a large upper portion of PDR decays to excited states \cite{Bassauer2020}. The suppression of the PDR's total $B_{mn}(E1\downarrow)$ by the multiplicity factor of three and another substantial reduction when excluding the cascade decays may lead to comparable total $B_{mn}(E1\downarrow)$ values for PDR and PDR$^{\ast}(2_1^+)$. Although it is difficult to accurately extract exclusive PDR decays to the ground state theoretically, it may be an interesting task for the future in the context of the Brink-Axel hypothesis (BAH) \cite{brink1955some,axel1962electric}. It was originally formulated for IVGDR, suggesting that its strength distribution is approximately independent of the specific nuclear state it is built on. The validity of BAH for PDR is actively debated \cite{isaak2013constraining,Bassauer2020,Sieja2023}, and there is evidence for its violation, but our analysis shows that comparing PDR and PDR$^{\ast}(2_1^+)$ should be done for a specific direction of the transitions and needs a clearer definition of PDR.  
%}

%
\begin{figure}
    \centering
    \includegraphics[width=0.9\columnwidth]{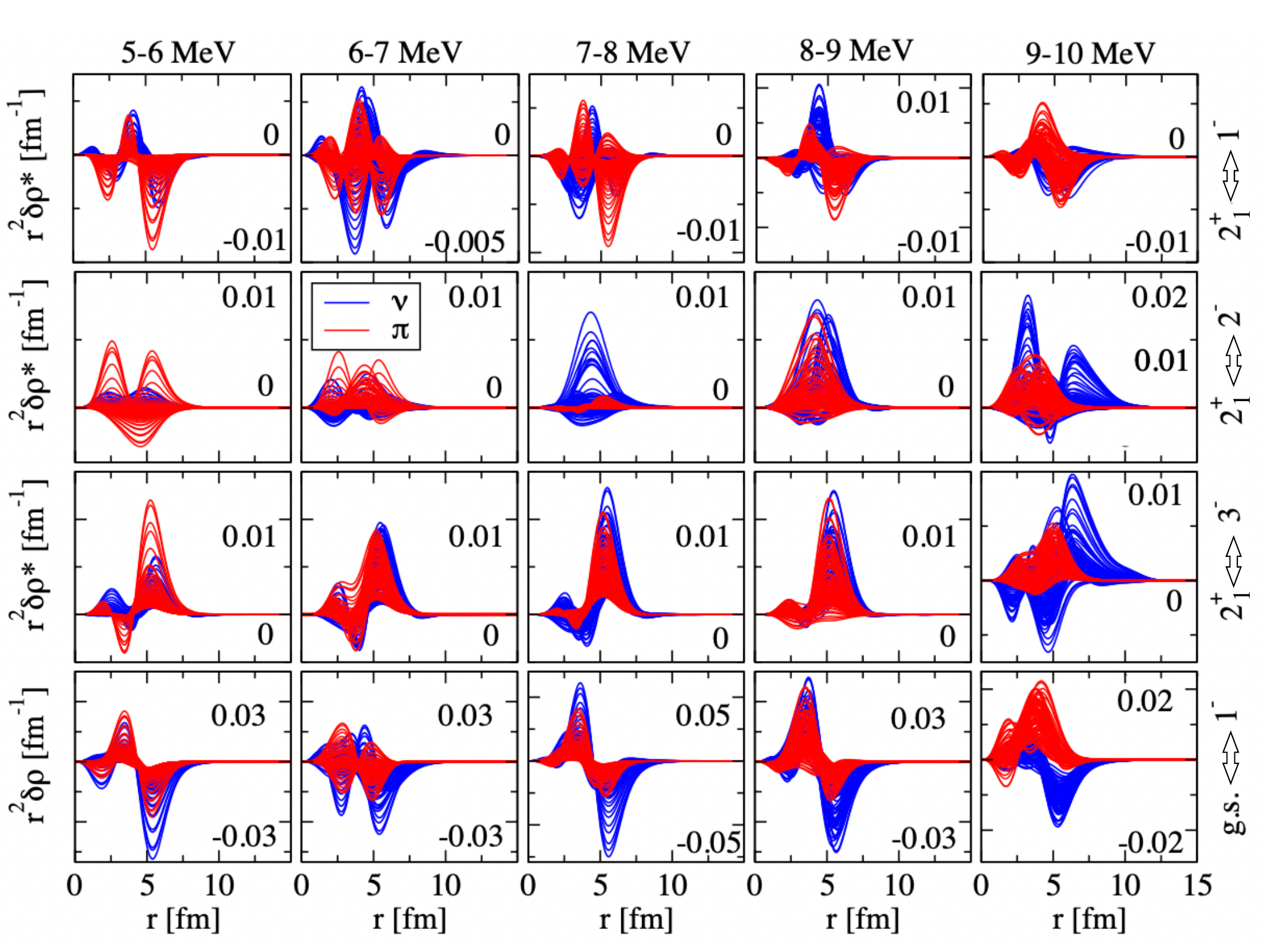}
    \caption{Evolution of the neutron ($\nu$) and proton ($\pi$) transition densities with energy plotted with the 10 keV energy step: the generalized transition densities $\delta\rho^{\ast}$ for the $J^{\pi} = (1,2,3)^-$ components of PDR$^{\ast}(2_1^+)$ (top three rows, respectively) and the regular transition densities $\delta\rho$ for the $J^{\pi} = 1^-$ ground-state-based PDR (bottom row).}
    \label{TrDen_all}
\end{figure}

Fig. \ref{TrDen_all} displays the palette of the relevant proton and neutron transition densities: the generalized transition densities for the $J^{\pi} = (1,2,3)^-$ components of PDR$^{\ast}(2_1^+)$ in comparison with the regular transition densities of the $J^{\pi} = 1^-$ ground-state-based PDR for the excited states below 10 MeV. The latter exhibits the behavior 
%of the neutron and proton transition densities 
typical for PDR: predominantly isoscalar (5-7 MeV), neutron-skin (7-9 MeV), and transitional (9-10 MeV) components can be distinguished, which qualitatively agrees with the case of $^{120}$Sn exemplified in Fig. \ref{fig0}(e). The dominance of coherent behavior of the $2_1^+ \leftrightarrow (2,3)^-$ generalized transition densities spans a larger energy interval up to $\sim$9 MeV before changing to the neutron-skin character, which is expressed in larger amplitudes and considerable spatial extension of the neutron components. This pattern is rooted in the predominantly isoscalar nature of the $2_1^+$ and $(2,3)^-$ states below this energy and in the emerging availability of loosely-bound neutron orbitals in the formation of the $(2,3)^-$ states above it. The $2_1^+ \leftrightarrow 1^-$ generalized transition densities start showing decoherence at 8-9 MeV and considerable proton dominance at 9-10 MeV, which is qualitatively consistent with the upper component of the ground-state-based PDR.

\section*{Summary and Outlook}

We formulated the concept and presented the first experimental evidence for the PDR of non-1$^-$ spin-parity built on a nuclear excited state.  
Employing the selective population of excited states in the neutron-rich $^{80}$Ge nucleus by the $\beta$ decay of the $J^{\pi} = 3^-$ isomer of $^{80m}$Ga  with a large Q$_{\beta}$ value, we accessed the $J >1$ components of this mode based on the lowest quadrupole excited state $2_1^+$, dubbed as PDR$^{\ast}(2_1^+)$. These components were identified experimentally as a group of $J^{\pi} = (2,3)^-$ excited states located between 5.6 MeV and the neutron emission threshold $S_n = 8.08$ MeV, with the two most prominent peaks at 7.840 and 7.996 MeV and substantial decay probabilities to the lowest quadrupole excited state.
% $2_1^+$. 
Microscopic calculations 
%within the REOM$^3$ approach and shell model analysis 
support the experimental findings in terms of the spin-parity assignment of these states and identifying them as PDR$^{\ast}(2_1^+)$. 
%In particular, 
Theory suggests that the two prominent states have $J^{\pi} = 3^-$ and are located at the onset of a pronounced neutron skin oscillation 
according to the transition density analysis, which also allows firm distinction of PDR$^{\ast}(2_1^+)$ from IVGDR. The computed strength distributions
justify the resonant character of the new mode concentrated in the octupole component.

This work opens an avenue for 
%experimental and theoretical 
further studies of the multicomponent PDR$^{\ast}$ built on non-$0^+$ excited states in neutron-rich nuclei, which plays a critical role in the r-process nucleosynthesis. In particular, experiments 
%search 
%for this phenomenon 
can be pursued with higher statistics and beyond $S_n$, while 
%the microscopic 
theory would benefit from further refinement in both interaction and many-body sectors for accurate interpretation of spectroscopic observations and reducing uncertainties in the astrophysical modeling.

\bibliography{80Ge-bibliography.bib}

\begin{thebibliography}{10}
\expandafter\ifx\csname url\endcsname\relax
  \def\url#1{\texttt{#1}}\fi
\expandafter\ifx\csname urlprefix\endcsname\relax\def\urlprefix{URL }\fi
\providecommand{\bibinfo}[2]{#2}
\providecommand{\eprint}[2][]{\url{#2}}

\bibitem{Pian2017}
\bibinfo{author}{Pian, E.} \emph{et~al.}
\newblock \bibinfo{title}{Spectroscopic identification of r-process
  nucleosynthesis in a double neutron star merger}.
\newblock \emph{\bibinfo{journal}{Nature}} \textbf{\bibinfo{volume}{551}},
  \bibinfo{pages}{67--70} (\bibinfo{year}{2017}).

\bibitem{Meszaros2019}
\bibinfo{author}{M\'esz\'aros, P.}, \bibinfo{author}{Fox, D.~B.},
  \bibinfo{author}{Hanna, C.} \& \bibinfo{author}{Murase, K.}
\newblock \bibinfo{title}{Multi-messenger astrophysics}.
\newblock \emph{\bibinfo{journal}{Nature Rev. Phys.}}
  \textbf{\bibinfo{volume}{1}}, \bibinfo{pages}{585--599}
  (\bibinfo{year}{2019}).

\bibitem{nasa}
\emph{\bibinfo{title}{https://www.nasa.gov/mission/chandra-x-ray-observatory/}}.

\bibitem{OezelTashenov2014}
\bibinfo{author}{\"Ozel-Tashenov, B.} \emph{et~al.}
\newblock \bibinfo{title}{Low-energy dipole strength in $^{112,120}${Sn}}.
\newblock \emph{\bibinfo{journal}{Phys. Rev. C}} \textbf{\bibinfo{volume}{90}},
  \bibinfo{pages}{024304} (\bibinfo{year}{2014}).

\bibitem{Bassauer2020}
\bibinfo{author}{Bassauer, S.} \emph{et~al.}
\newblock \bibinfo{title}{Electric and magnetic dipole strength in
  $^{112,114,116,118,120,124}${Sn}}.
\newblock \emph{\bibinfo{journal}{Phys. Rev. C}}
  \textbf{\bibinfo{volume}{102}}, \bibinfo{pages}{034327}
  (\bibinfo{year}{2020}).
\newblock
  \urlprefix\url{https://journals.aps.org/prc/abstract/10.1103/PhysRevC.102.034327}.

\bibitem{bracco2019isoscalar}
\bibinfo{author}{Bracco, A.}, \bibinfo{author}{Lanza, E.} \&
  \bibinfo{author}{Tamii, A.}
\newblock \bibinfo{title}{Isoscalar and isovector dipole excitations: Nuclear
  properties from low-lying states and from the isovector giant dipole
  resonance}.
\newblock \emph{\bibinfo{journal}{Prog. Part. Nucl. Phys.}}
  \textbf{\bibinfo{volume}{106}}, \bibinfo{pages}{360--433}
  (\bibinfo{year}{2019}).

\bibitem{harakeh2001giant}
\bibinfo{author}{Harakeh, M.~N.} \& \bibinfo{author}{{van der Woude}, A.}
\newblock \emph{\bibinfo{title}{Giant Resonances: fundamental high-frequency
  modes of nuclear excitation}} (\bibinfo{publisher}{Oxford University Press},
  \bibinfo{address}{Oxford}, \bibinfo{year}{2001}).

\bibitem{bartholomew1961neutron}
\bibinfo{author}{Bartholomew, G.}
\newblock \bibinfo{title}{Neutron capture gamma rays}.
\newblock \emph{\bibinfo{journal}{Annu. Rev. Nucl. Sci.}}
  \textbf{\bibinfo{volume}{11}}, \bibinfo{pages}{259--302}
  (\bibinfo{year}{1961}).

\bibitem{rossi2013measurement}
\bibinfo{author}{Rossi, D.} \emph{et~al.}
\newblock \bibinfo{title}{Measurement of the dipole polarizability of the
  unstable neutron-rich nucleus $^{68}${Ni}}.
\newblock \emph{\bibinfo{journal}{Phys. Rev. Lett.}}
  \textbf{\bibinfo{volume}{111}}, \bibinfo{pages}{242503}
  (\bibinfo{year}{2013}).
\newblock
  \urlprefix\url{https://journals.aps.org/prl/abstract/10.1103/PhysRevLett.111.242503}.

\bibitem{mohan1971three}
\bibinfo{author}{Mohan, R.}, \bibinfo{author}{Danos, M.} \&
  \bibinfo{author}{Biedenharn, L.}
\newblock \bibinfo{title}{Three-fluid hydrodynamical model of nuclei}.
\newblock \emph{\bibinfo{journal}{Phys. Rev. C}} \textbf{\bibinfo{volume}{3}},
  \bibinfo{pages}{1740} (\bibinfo{year}{1971}).

\bibitem{Vretenar2001}
\bibinfo{author}{Vretenar, D.}, \bibinfo{author}{Paar, N.},
  \bibinfo{author}{Ring, P.} \& \bibinfo{author}{Lalazissis, G.~A.}
\newblock \bibinfo{title}{Pygmy dipole resonances in the relativistic random
  phase approximation}.
\newblock \emph{\bibinfo{journal}{Phys. Rev. C}} \textbf{\bibinfo{volume}{63}},
  \bibinfo{pages}{047301} (\bibinfo{year}{2001}).

\bibitem{savran2013experimental}
\bibinfo{author}{Savran, D.}, \bibinfo{author}{Aumann, T.} \&
  \bibinfo{author}{Zilges, A.}
\newblock \bibinfo{title}{Experimental studies of the pygmy dipole resonance}.
\newblock \emph{\bibinfo{journal}{Prog. Part. Nucl. Phys.}}
  \textbf{\bibinfo{volume}{70}}, \bibinfo{pages}{210--245}
  (\bibinfo{year}{2013}).

\bibitem{piekarewicz2011pygmy}
\bibinfo{author}{Piekarewicz, J.}
\newblock \bibinfo{title}{Pygmy resonances and neutron skins}.
\newblock \emph{\bibinfo{journal}{Phys. Rev. C}} \textbf{\bibinfo{volume}{83}},
  \bibinfo{pages}{034319} (\bibinfo{year}{2011}).

\bibitem{klimkiewicz2007nuclear}
\bibinfo{author}{Klimkiewicz, A.} \emph{et~al.}
\newblock \bibinfo{title}{Nuclear symmetry energy and neutron skins derived
  from pygmy dipole resonances}.
\newblock \emph{\bibinfo{journal}{Phys. Rev. C}} \textbf{\bibinfo{volume}{76}},
  \bibinfo{pages}{051603(R)} (\bibinfo{year}{2007}).

\bibitem{Roca2015}
\bibinfo{author}{Roca-Maza, X.} \emph{et~al.}
\newblock \bibinfo{title}{Neutron skin thickness from the measured electric
  dipole polarizability in $^{68}\text{Ni}$, $^{120}\text{Sn}$, and
  $^{208}\text{Pb}$}.
\newblock \emph{\bibinfo{journal}{Physical Review C}}
  \textbf{\bibinfo{volume}{92}}, \bibinfo{pages}{064304}
  (\bibinfo{year}{2015}).
\newblock \urlprefix\url{https://link.aps.org/doi/10.1103/PhysRevC.92.064304}.

\bibitem{fattoyev2018neutron}
\bibinfo{author}{Fattoyev, F.~J.}, \bibinfo{author}{Piekarewicz, J.} \&
  \bibinfo{author}{Horowitz, C.~J.}
\newblock \bibinfo{title}{Neutron skins and neutron stars in the multimessenger
  era}.
\newblock \emph{\bibinfo{journal}{Phys. Rev. Lett.}}
  \textbf{\bibinfo{volume}{120}}, \bibinfo{pages}{172702}
  (\bibinfo{year}{2018}).

\bibitem{goriely1998radiative}
\bibinfo{author}{Goriely, S.}
\newblock \bibinfo{title}{Radiative neutron captures by neutron-rich nuclei and
  the r-process nucleosynthesis}.
\newblock \emph{\bibinfo{journal}{Phys. Lett. B}}
  \textbf{\bibinfo{volume}{436}}, \bibinfo{pages}{10--18}
  (\bibinfo{year}{1998}).

\bibitem{spieker2020accessing}
\bibinfo{author}{Spieker, M.} \emph{et~al.}
\newblock \bibinfo{title}{Accessing the single-particle structure of the pygmy
  dipole resonance in $^{208}${Pb}}.
\newblock \emph{\bibinfo{journal}{Phys. Rev. Lett.}}
  \textbf{\bibinfo{volume}{125}}, \bibinfo{pages}{102503}
  (\bibinfo{year}{2020}).

\bibitem{paar2007exotic}
\bibinfo{author}{Paar, N.}, \bibinfo{author}{Vretenar, D.},
  \bibinfo{author}{Khan, E.} \& \bibinfo{author}{Colò, G.}
\newblock \bibinfo{title}{Exotic modes of excitation in atomic nuclei far from
  stability}.
\newblock \emph{\bibinfo{journal}{Rep. Prog. Phys.}}
  \textbf{\bibinfo{volume}{70}}, \bibinfo{pages}{691} (\bibinfo{year}{2007}).

\bibitem{endres2009splitting}
\bibinfo{author}{Endres, J.} \emph{et~al.}
\newblock \bibinfo{title}{Splitting of the pygmy dipole resonance in
  $^{138}${Ba} and $^{140}${Ce} observed in the ($\alpha$, $\alpha^{\prime}$
  $\gamma$) reaction}.
\newblock \emph{\bibinfo{journal}{Phys. Rev. C}} \textbf{\bibinfo{volume}{80}},
  \bibinfo{pages}{034302} (\bibinfo{year}{2009}).

\bibitem{savran2006nature}
\bibinfo{author}{Savran, D.} \emph{et~al.}
\newblock \bibinfo{title}{Nature of the pygmy dipole resonance in $^{140}${Ce}
  studied in ($\alpha$, $\alpha^{\prime}$ $\gamma$) experiments}.
\newblock \emph{\bibinfo{journal}{Phys. Rev. Lett.}}
  \textbf{\bibinfo{volume}{97}}, \bibinfo{pages}{172502}
  (\bibinfo{year}{2006}).

\bibitem{endres2010isospin}
\bibinfo{author}{Endres, J.} \emph{et~al.}
\newblock \bibinfo{title}{Isospin character of the pygmy dipole resonance in
  $^{124}${Sn}}.
\newblock \emph{\bibinfo{journal}{Phys. Rev. Lett.}}
  \textbf{\bibinfo{volume}{105}}, \bibinfo{pages}{212503}
  (\bibinfo{year}{2010}).

\bibitem{paar2009isoscalar}
\bibinfo{author}{Paar, N.}, \bibinfo{author}{Niu, Y.},
  \bibinfo{author}{Vretenar, D.} \& \bibinfo{author}{Meng, J.}
\newblock \bibinfo{title}{Isoscalar and isovector splitting of pygmy dipole
  structures}.
\newblock \emph{\bibinfo{journal}{Phys. Rev. Lett.}}
  \textbf{\bibinfo{volume}{103}}, \bibinfo{pages}{032502}
  (\bibinfo{year}{2009}).

\bibitem{lanza2014dipole}
\bibinfo{author}{Lanza, E.~G.}, \bibinfo{author}{Vitturi, A.},
  \bibinfo{author}{Litvinova, E.} \& \bibinfo{author}{Savran, D.}
\newblock \bibinfo{title}{Dipole excitations via isoscalar probes: The
  splitting of the pygmy dipole resonance in $^{124}${Sn}}.
\newblock \emph{\bibinfo{journal}{Phys. Rev. C}} \textbf{\bibinfo{volume}{89}},
  \bibinfo{pages}{041601} (\bibinfo{year}{2014}).

\bibitem{Angell2012}
\bibinfo{author}{Angell, C.~T.} \emph{et~al.}
\newblock \bibinfo{title}{Evidence for radiative coupling of the pygmy dipole
  resonance to excited states}.
\newblock \emph{\bibinfo{journal}{Phys. Rev. C}} \textbf{\bibinfo{volume}{86}},
  \bibinfo{pages}{051302} (\bibinfo{year}{2012}).
\newblock \urlprefix\url{https://link.aps.org/doi/10.1103/PhysRevC.86.051302}.

\bibitem{scheck2013decay}
\bibinfo{author}{Scheck, M.} \emph{et~al.}
\newblock \bibinfo{title}{Decay pattern of the pygmy dipole resonance in
  $^{60}${Ni}}.
\newblock \emph{\bibinfo{journal}{Phys. Rev. C}} \textbf{\bibinfo{volume}{87}},
  \bibinfo{pages}{051304(R)} (\bibinfo{year}{2013}).

\bibitem{isaak2013constraining}
\bibinfo{author}{Isaak, J.} \emph{et~al.}
\newblock \bibinfo{title}{Constraining nuclear photon strength functions by the
  decay properties of photo-excited states}.
\newblock \emph{\bibinfo{journal}{Phys. Lett. B}}
  \textbf{\bibinfo{volume}{727}}, \bibinfo{pages}{361--365}
  (\bibinfo{year}{2013}).

\bibitem{loher2016decay}
\bibinfo{author}{L{\"o}her, B.} \emph{et~al.}
\newblock \bibinfo{title}{The decay pattern of the pygmy dipole resonance of
  $^{140}${Ce}}.
\newblock \emph{\bibinfo{journal}{Phys. Lett. B}}
  \textbf{\bibinfo{volume}{756}}, \bibinfo{pages}{72--76}
  (\bibinfo{year}{2016}).

\bibitem{Markova2025}
\bibinfo{author}{Markova, M.}, \bibinfo{author}{von Neumann-Cosel, P.} \&
  \bibinfo{author}{Litvinova, E.}
\newblock \bibinfo{title}{Systematics of the low-energy electric dipole
  strength in the sn isotopic chain}.
\newblock \emph{\bibinfo{journal}{Phys. Lett. B}}
  \textbf{\bibinfo{volume}{860}}, \bibinfo{pages}{139216}
  (\bibinfo{year}{2025}).

\bibitem{PhysRevC.100.064320}
\bibinfo{author}{Litvinova, E.} \& \bibinfo{author}{Schuck, P.}
\newblock \bibinfo{title}{Toward an accurate strongly coupled many-body theory
  within the equation-of-motion framework}.
\newblock \emph{\bibinfo{journal}{Phys. Rev. C}}
  \textbf{\bibinfo{volume}{100}}, \bibinfo{pages}{064320}
  (\bibinfo{year}{2019}).
\newblock \urlprefix\url{https://link.aps.org/doi/10.1103/PhysRevC.100.064320}.

\bibitem{novak2024}
\bibinfo{author}{Novak, J.}, \bibinfo{author}{Hlatshwayo, M.~Q.} \&
  \bibinfo{author}{Litvinova, E.}
\newblock \bibinfo{title}{Response of strongly coupled fermions on classical
  and quantum computers} (\bibinfo{year}{2024}).
\newblock \urlprefix\url{https://arxiv.org/abs/2405.02255}.
\newblock \bibinfo{note}{ArXiv:2405.02255}, \eprint{2405.02255}.

\bibitem{Tonchev2010}
\bibinfo{author}{Tonchev, A.~P.} \emph{et~al.}
\newblock \bibinfo{title}{Spectral structure of the pygmy dipole resonance}.
\newblock \emph{\bibinfo{journal}{Phys. Rev. Lett.}}
  \textbf{\bibinfo{volume}{104}}, \bibinfo{pages}{072501}
  (\bibinfo{year}{2010}).

\bibitem{Derya2013}
\bibinfo{author}{Derya, V.} \emph{et~al.}
\newblock \bibinfo{title}{Study of the pygmy dipole resonance in $^{94}${Mo}
  using the ($\alpha$, $\alpha '\gamma$) coincidence technique}.
\newblock \emph{\bibinfo{journal}{Nucl. Phys. A}}
  \textbf{\bibinfo{volume}{906}}, \bibinfo{pages}{94--107}
  (\bibinfo{year}{2013}).

\bibitem{mashtakov2021structure}
\bibinfo{author}{Mashtakov, K.~R.} \emph{et~al.}
\newblock \bibinfo{title}{Structure of high-lying levels populated in the
  $^{96}${Y} $\rightarrow$ $^{96}${Zr} $\beta$ decay}.
\newblock \emph{\bibinfo{journal}{Phys. Lett. B}} \bibinfo{pages}{136569}
  (\bibinfo{year}{2021}).

\bibitem{bohr1998nuclear}
\bibinfo{author}{Bohr, A.} \& \bibinfo{author}{Mottelson, B.~R.}
\newblock \emph{\bibinfo{title}{Nuclear Structure (in 2 volumes)}}
  (\bibinfo{publisher}{World Scientific Publishing Company},
  \bibinfo{address}{Singapore$\cdot$New Jersey$\cdot$London$\cdot$HongKong},
  \bibinfo{year}{1998}).

\bibitem{Adrich2005}
\bibinfo{author}{Adrich, P.} \emph{et~al.}
\newblock \bibinfo{title}{Evidence for pygmy and giant dipole resonances in
  {Sn}-130 and {Sn}-132}.
\newblock \emph{\bibinfo{journal}{Phys. Rev. Lett.}}
  \textbf{\bibinfo{volume}{95}}, \bibinfo{pages}{132501}
  (\bibinfo{year}{2005}).

\bibitem{wieland2009search}
\bibinfo{author}{Wieland, O.} \emph{et~al.}
\newblock \bibinfo{title}{Search for the pygmy dipole resonance in $^{68}${Ni}
  at 600 mev/nucleon}.
\newblock \emph{\bibinfo{journal}{Phys. Rev. Lett.}}
  \textbf{\bibinfo{volume}{102}}, \bibinfo{pages}{092502}
  (\bibinfo{year}{2009}).

\bibitem{Wieland2018}
\bibinfo{author}{Wieland, O.} \emph{et~al.}
\newblock \bibinfo{title}{Low-lying dipole response in the unstable $^{70}${Ni}
  nucleus}.
\newblock \emph{\bibinfo{journal}{Phys. Rev. C}} \textbf{\bibinfo{volume}{98}},
  \bibinfo{pages}{064313} (\bibinfo{year}{2018}).

\bibitem{zilges2022photonuclear}
\bibinfo{author}{Zilges, A.}, \bibinfo{author}{Balabanski, D.},
  \bibinfo{author}{Isaak, J.} \& \bibinfo{author}{Pietralla, N.}
\newblock \bibinfo{title}{Photonuclear reactions — from basic research to
  applications}.
\newblock \emph{\bibinfo{journal}{Prog. Part. Nucl. Phys.}}
  \textbf{\bibinfo{volume}{122}}, \bibinfo{pages}{103903}
  (\bibinfo{year}{2022}).

\bibitem{tamii2011complete}
\bibinfo{author}{Tamii, A.} \emph{et~al.}
\newblock \bibinfo{title}{Complete electric dipole response and the neutron
  skin in $^{208}${Pb}}.
\newblock \emph{\bibinfo{journal}{Phys. Rev. Lett.}}
  \textbf{\bibinfo{volume}{107}}, \bibinfo{pages}{062502}
  (\bibinfo{year}{2011}).

\bibitem{crespi2014isospin}
\bibinfo{author}{Crespi, F.} \emph{et~al.}
\newblock \bibinfo{title}{Isospin character of low-lying pygmy dipole states in
  $^{208}${Pb} via inelastic scattering of $^{17}${O} ions}.
\newblock \emph{\bibinfo{journal}{Phys. Rev. Lett.}}
  \textbf{\bibinfo{volume}{113}}, \bibinfo{pages}{012501}
  (\bibinfo{year}{2014}).

\bibitem{weinert2021microscopic}
\bibinfo{author}{Weinert, M.} \emph{et~al.}
\newblock \bibinfo{title}{Microscopic structure of the low-energy electric
  dipole response of $^{120}${Sn}}.
\newblock \emph{\bibinfo{journal}{Phys. Rev. Lett.}}
  \textbf{\bibinfo{volume}{127}}, \bibinfo{pages}{242501}
  (\bibinfo{year}{2021}).

\bibitem{scheck2016investigating}
\bibinfo{author}{Scheck, M.} \emph{et~al.}
\newblock \bibinfo{title}{Investigating the pygmy dipole resonance using
  $\beta$ decay}.
\newblock \emph{\bibinfo{journal}{Phys. Rev. Lett.}}
  \textbf{\bibinfo{volume}{116}}, \bibinfo{pages}{132501}
  (\bibinfo{year}{2016}).

\bibitem{gottardo2017unexpected}
\bibinfo{author}{Gottardo, A.} \emph{et~al.}
\newblock \bibinfo{title}{Unexpected high-energy $\gamma$ emission from
  decaying exotic nuclei}.
\newblock \emph{\bibinfo{journal}{Phys. Lett. B}}
  \textbf{\bibinfo{volume}{772}}, \bibinfo{pages}{359--362}
  (\bibinfo{year}{2017}).

\bibitem{ibrahim2007alto}
\bibinfo{author}{Ibrahim, F.} \emph{et~al.}
\newblock \bibinfo{title}{The {ALTO} facility at {IPN} {Orsay} and study of
  neutron rich nuclei in the vincinity of $^{78}${Ni}}.
\newblock \emph{\bibinfo{journal}{Nucl. Phys. A}}
  \textbf{\bibinfo{volume}{787}}, \bibinfo{pages}{110--117}
  (\bibinfo{year}{2007}).

\bibitem{cheal2010discovery}
\bibinfo{author}{Cheal, B.} \emph{et~al.}
\newblock \bibinfo{title}{Discovery of a long-lived low-lying isomeric state in
  $^{80}${Ga}}.
\newblock \emph{\bibinfo{journal}{Phys. Rev. C}} \textbf{\bibinfo{volume}{82}},
  \bibinfo{pages}{051302(R)} (\bibinfo{year}{2010}).

\bibitem{etile2015low}
\bibinfo{author}{Etil{\'e}, A.} \emph{et~al.}
\newblock \bibinfo{title}{Low-lying intruder and tensor-driven structures in
  $^{82}${As} revealed by $\beta$ decay at a new movable-tape-based
  experimental setup}.
\newblock \emph{\bibinfo{journal}{Phys. Rev. C}} \textbf{\bibinfo{volume}{91}},
  \bibinfo{pages}{064317} (\bibinfo{year}{2015}).

\bibitem{ciemala2009measurements}
\bibinfo{author}{Ciema{\l}a, M.} \emph{et~al.}
\newblock \bibinfo{title}{Measurements of high-energy $\gamma$-rays with
  {LaBr}$_3$:{Ce} detectors}.
\newblock \emph{\bibinfo{journal}{Nucl. Instrum. Methods Phys. Res. A}}
  \textbf{\bibinfo{volume}{608}}, \bibinfo{pages}{76--79}
  (\bibinfo{year}{2009}).

\bibitem{ghosh2016characterization}
\bibinfo{author}{Ghosh, C.} \emph{et~al.}
\newblock \bibinfo{title}{Characterization of paris {LaBr}$_3${(Ce)}-{NaI(Tl)}
  phoswich detectors up to {E}$_{\gamma}$ ~ 22 mev}.
\newblock \emph{\bibinfo{journal}{J. Instrum.}} \textbf{\bibinfo{volume}{11}},
  \bibinfo{pages}{P05023} (\bibinfo{year}{2016}).

\bibitem{wang2017ame2016}
\bibinfo{author}{Wang, M.} \emph{et~al.}
\newblock \bibinfo{title}{The {AME2016} atomic mass evaluation (ii). tables,
  graphs and references}.
\newblock \emph{\bibinfo{journal}{Chin. Phys. C}}
  \textbf{\bibinfo{volume}{41}}, \bibinfo{pages}{030003}
  (\bibinfo{year}{2017}).

\bibitem{ren2022thesis}
\bibinfo{author}{Li, R.}
\newblock \emph{\bibinfo{title}{{First attempt toward a quasi-Pandemonium free
  $\beta$-delayed spectroscopy of $^{80}$Ge using PARIS at ALTO}}}.
\newblock \bibinfo{type}{{PhD} {Thesis}}, \bibinfo{school}{{Universit{\'e}
  Paris-Saclay}} (\bibinfo{year}{2022}).
\newblock \urlprefix\url{https://theses.hal.science/tel-04172231}.

\bibitem{Li2025a}
\bibinfo{author}{Li, R.} \emph{et~al.}
\newblock \bibinfo{title}{Simultaneous impacts of nuclear shell structure and
  collectivity on beta decay: Evidence from 80-{Ga}}.
\newblock \emph{\bibinfo{journal}{Phys. Rev. C}}
  \textbf{\bibinfo{volume}{111}}, \bibinfo{pages}{034303}
  (\bibinfo{year}{2025}).

\bibitem{li_hal05035090}
\bibinfo{author}{Li, R.} \emph{et~al.}
\newblock \bibinfo{title}{{$\beta$-delayed spectroscopy of $^{80}$Ge$_{48}$}}.
\newblock \emph{\bibinfo{journal}{arXiv.2504.02156}}  (\bibinfo{year}{2025}).
\newblock \urlprefix\url{https://hal.science/hal-05035090}.

\bibitem{verney2013structure}
\bibinfo{author}{Verney, D.} \emph{et~al.}
\newblock \bibinfo{title}{Structure of $^{80}${Ge} revealed by the $\beta$
  decay of isomeric states in $^{80}${Ga}: Triaxiality in the vicinity of
  $^{78}${Ni}}.
\newblock \emph{\bibinfo{journal}{Phys. Rev. C}} \textbf{\bibinfo{volume}{87}},
  \bibinfo{pages}{054307} (\bibinfo{year}{2013}).

\bibitem{singh1998review}
\bibinfo{author}{Singh, B.}, \bibinfo{author}{Rodriguez, J.},
  \bibinfo{author}{Wong, S.} \& \bibinfo{author}{Tuli, J.}
\newblock \bibinfo{title}{Review of logft values in $\beta$ decay}.
\newblock \emph{\bibinfo{journal}{Nucl. Data Sheets}}
  \textbf{\bibinfo{volume}{84}}, \bibinfo{pages}{487--563}
  (\bibinfo{year}{1998}).

\bibitem{PARIS}
\bibinfo{author}{Bracco, A.} \emph{et~al.}
\newblock \bibinfo{title}{{PARIS} collaboration website}.
\newblock
  \bibinfo{howpublished}{\url{http://paris.ifj.edu.pl/index.php?lng=en}}
  (\bibinfo{year}{2022}).

\bibitem{FASTER}
\bibinfo{author}{{LPC CAEN}}.
\newblock \bibinfo{title}{{Fast} {Acquisition} {SysTem} for {nuclEar}
  {Research} (faster)}.
\newblock \bibinfo{howpublished}{\url{http://faster.in2p3.fr/}}
  (\bibinfo{year}{2021}).

\bibitem{farooq2017probing}
\bibinfo{author}{Farooq-Smith, G.~J.} \emph{et~al.}
\newblock \bibinfo{title}{Probing the {Ga}$_{31}$ ground-state properties in
  the region near {Z=28} with high-resolution laser spectroscopy}.
\newblock \emph{\bibinfo{journal}{Phys. Rev. C}} \textbf{\bibinfo{volume}{96}},
  \bibinfo{pages}{044324} (\bibinfo{year}{2017}).

\bibitem{iwasaki2008persistence}
\bibinfo{author}{Iwasaki, H.} \emph{et~al.}
\newblock \bibinfo{title}{Persistence of the {N=50} shell closure in the
  neutron-rich isotope $^{80}${Ge}}.
\newblock \emph{\bibinfo{journal}{Phys. Rev. C}} \textbf{\bibinfo{volume}{78}},
  \bibinfo{pages}{021304(R)} (\bibinfo{year}{2008}).

\bibitem{padilla2005b}
\bibinfo{author}{Padilla-Rodal, E.} \emph{et~al.}
\newblock \bibinfo{title}{{B(E2)}$\uparrow$ measurements for radioactive
  neutron-rich {Ge} isotopes: Reaching the {N=50} closed shell}.
\newblock \emph{\bibinfo{journal}{Phys. Rev. Lett.}}
  \textbf{\bibinfo{volume}{94}}, \bibinfo{pages}{122501}
  (\bibinfo{year}{2005}).

\bibitem{PLANT1998607}
\bibinfo{author}{Plant, R.~S.} \& \bibinfo{author}{Birse, M.~C.}
\newblock \bibinfo{title}{Meson properties in an extended non-local njl model}.
\newblock \emph{\bibinfo{journal}{Nuclear Physics A}}
  \textbf{\bibinfo{volume}{628}}, \bibinfo{pages}{607--644}
  (\bibinfo{year}{1998}).
\newblock
  \urlprefix\url{https://www.sciencedirect.com/science/article/pii/S0375947497006350}.

\bibitem{PhysRevC.78.014312}
\bibinfo{author}{Litvinova, E.}, \bibinfo{author}{Ring, P.} \&
  \bibinfo{author}{Tselyaev, V.}
\newblock \bibinfo{title}{Relativistic quasiparticle time blocking
  approximation: Dipole response of open-shell nuclei}.
\newblock \emph{\bibinfo{journal}{Phys. Rev. C}} \textbf{\bibinfo{volume}{78}},
  \bibinfo{pages}{014312} (\bibinfo{year}{2008}).
\newblock \urlprefix\url{https://link.aps.org/doi/10.1103/PhysRevC.78.014312}.

\bibitem{PhysRevC.88.054301}
\bibinfo{author}{Tselyaev, V.~I.}
\newblock \bibinfo{title}{Subtraction method and stability condition in
  extended random-phase approximation theories}.
\newblock \emph{\bibinfo{journal}{Phys. Rev. C}} \textbf{\bibinfo{volume}{88}},
  \bibinfo{pages}{054301} (\bibinfo{year}{2013}).
\newblock \urlprefix\url{https://link.aps.org/doi/10.1103/PhysRevC.88.054301}.

\bibitem{PhysRevC.106.064316}
\bibinfo{author}{Litvinova, E.} \& \bibinfo{author}{Zhang, Y.}
\newblock \bibinfo{title}{Microscopic response theory for strongly coupled
  superfluid fermionic systems}.
\newblock \emph{\bibinfo{journal}{Phys. Rev. C}}
  \textbf{\bibinfo{volume}{106}}, \bibinfo{pages}{064316}
  (\bibinfo{year}{2022}).
\newblock \urlprefix\url{https://link.aps.org/doi/10.1103/PhysRevC.106.064316}.

\bibitem{Paar2003}
\bibinfo{author}{Paar, N.}, \bibinfo{author}{Ring, P.},
  \bibinfo{author}{Nik{\v{s}}i{\'c}, T.} \& \bibinfo{author}{Vretenar, D.}
\newblock \bibinfo{title}{Quasiparticle random phase approximation based on the
  relativistic hartree-bogoliubov model}.
\newblock \emph{\bibinfo{journal}{Phys. Rev. C}} \textbf{\bibinfo{volume}{67}},
  \bibinfo{pages}{034312} (\bibinfo{year}{2003}).

\bibitem{LALAZISSIS200936}
\bibinfo{author}{Lalazissis, G.~A.} \emph{et~al.}
\newblock \bibinfo{title}{The effective force {NL}3 revisited}.
\newblock \emph{\bibinfo{journal}{Physics Letters B}}
  \textbf{\bibinfo{volume}{671}}, \bibinfo{pages}{36--41}
  (\bibinfo{year}{2009}).
\newblock
  \urlprefix\url{https://www.sciencedirect.com/science/article/pii/S0370269308014500}.

\bibitem{RIPL3}
\emph{\bibinfo{title}{https://www-nds.iaea.org/RIPL-3/}}.

\bibitem{BERTULANI1999139}
\bibinfo{author}{Bertulani, C.~A.} \& \bibinfo{author}{Ponomarev, V.~Y.}
\newblock \bibinfo{title}{Microscopic studies on two-phonon giant resonances}.
\newblock \emph{\bibinfo{journal}{Physics Reports}}
  \textbf{\bibinfo{volume}{321}}, \bibinfo{pages}{139--251}
  (\bibinfo{year}{1999}).
\newblock
  \urlprefix\url{https://www.sciencedirect.com/science/article/pii/S0370157399000381}.

\bibitem{RS80}
\bibinfo{author}{Ring, P.} \& \bibinfo{author}{Schuck, P.}
\newblock \emph{\bibinfo{title}{The Nuclear Many-Body Problems}}
  (\bibinfo{publisher}{Springer}, \bibinfo{address}{Verlag Berlin Heidelberg},
  \bibinfo{year}{1980}).

\bibitem{Edmonds}
\bibinfo{author}{Edmonds, A.~R.}
\newblock \emph{\bibinfo{title}{Angular Momentum in Quantum Mechanics}}
  (\bibinfo{publisher}{Princeton University Press, Princeton},
  \bibinfo{year}{1957}).

\bibitem{PhysRevC.36.380}
\bibinfo{author}{Bouyssy, A.}, \bibinfo{author}{Mathiot, J.-F.},
  \bibinfo{author}{Van~Giai, N.} \& \bibinfo{author}{Marcos, S.}
\newblock \bibinfo{title}{Relativistic description of nuclear systems in the
  hartree-fock approximation}.
\newblock \emph{\bibinfo{journal}{Phys. Rev. C}} \textbf{\bibinfo{volume}{36}},
  \bibinfo{pages}{380--401} (\bibinfo{year}{1987}).
\newblock \urlprefix\url{https://link.aps.org/doi/10.1103/PhysRevC.36.380}.

\bibitem{Ma2002}
\bibinfo{author}{Cao, L.-g.} \& \bibinfo{author}{Ma, Z.-y.}
\newblock \bibinfo{title}{Exploration of resonant continuum and giant resonance
  in the relativistic approach}.
\newblock \emph{\bibinfo{journal}{Physical Review C}}
  \textbf{\bibinfo{volume}{66}}, \bibinfo{pages}{024311}
  (\bibinfo{year}{2002}).

\bibitem{Alhassid1982}
\bibinfo{author}{Alhassid, Y.}, \bibinfo{author}{Gai, M.} \&
  \bibinfo{author}{Bertsch, G.~F.}
\newblock \bibinfo{title}{Radiative width of molecular-cluster states}.
\newblock \emph{\bibinfo{journal}{Physical Review Letters}}
  \textbf{\bibinfo{volume}{49}}, \bibinfo{pages}{1482--1485}
  (\bibinfo{year}{1982}).

\bibitem{Rugari1993}
\bibinfo{author}{Rugari, S.~L.} \emph{et~al.}
\newblock \bibinfo{title}{Broken reflection symmetry in {Xe}-114}.
\newblock \emph{\bibinfo{journal}{Physical Review C}}
  \textbf{\bibinfo{volume}{48}}, \bibinfo{pages}{2078--2081}
  (\bibinfo{year}{1993}).

\bibitem{Butler2019}
\bibinfo{author}{Butler, P.~A.} \emph{et~al.}
\newblock \bibinfo{title}{The observation of vibrating pear-shapes in radon
  nuclei}.
\newblock \emph{\bibinfo{journal}{Nature Communications}}
  \textbf{\bibinfo{volume}{10}}, \bibinfo{pages}{2473} (\bibinfo{year}{2019}).

\bibitem{brink1955some}
\bibinfo{author}{Brink, D.}
\newblock \emph{\bibinfo{title}{Some aspects of the interaction of light with
  matter}}.
\newblock Ph.D. thesis,
  \bibinfo{school}{\href{https://ora.ox.ac.uk/objects/uuid:334ec4a3-8a89-42aa-93f4-2e54d070ee09}{University
  of Oxford}} (\bibinfo{year}{1955}).

\bibitem{axel1962electric}
\bibinfo{author}{Axel, P.}
\newblock \bibinfo{title}{Electric dipole ground-state transition width
  strength function and {7-MeV} photon interactions}.
\newblock \emph{\bibinfo{journal}{Phys. Rev.}} \textbf{\bibinfo{volume}{126}},
  \bibinfo{pages}{671} (\bibinfo{year}{1962}).

\bibitem{Sieja2023}
\bibinfo{author}{Sieja, K.}
\newblock \bibinfo{title}{Brink-axel hypothesis in the pygmy-dipole resonance
  region}.
\newblock \emph{\bibinfo{journal}{Eur. Phys. J. A}}
  \textbf{\bibinfo{volume}{59}}, \bibinfo{pages}{147} (\bibinfo{year}{2023}).

\end{thebibliography}

\section*{Data availability}

Raw data are available on reasonable request.

\section*{Code availability}

Codes used for data processing and analysis are available upon reasonable request.

\section*{Acknowledgments}

The authors gratefully acknowledge the work of the ALTO staff for the excellent operation of the ISOL source and data acquisition. R. Li acknowledges support by China Scholarship Council under Grant No.201804910509 and KU Leuven postdoctoral fellow scholarship. The work of E. Litvinova was supported by the GANIL Visitor Program and US-NSF Grants PHY-2209376 and PHY-2515056. A. Kankainen and L. A. Ayoubi have received funding from the European Union's Horizon 2020 research and innovation program under grant agreement NO. 771036 (ERC CoG MAIDEN). S. Ebata was supported by the Leading Initiative for Excellent Young Researchers, MEXT, Japan. Use of the PARIS modular array from the PARIS collaboration and Ge detectors from the French-UK IN2P3-STFC Gamma Loan Pool is acknowledged. The authors acknowledge enlightening discussions with K. Sieja, M. Lewitowicz, J. Piot, and D. Ackermann, as well as the theoretical work on the quasiparticle-phonon model calculations by V. Yu. Ponomarev, which helped interpret experimental data and identify low-lying states in $^{80}$Ge during the early stages of manuscript preparation. 

\section*{Author contributions}

R. L. initiated the concept of PDR, built on nuclear quadrupole collectivity, participated in the experiment, analyzed data, and wrote the initial version of the manuscript. E. L. developed the REOM$^3$ method, performed theoretical calculations, contributed to the interpretation of results, and revised the article. M. N. H. contributed to the theoretical discussion, interpretation of results, manuscript preparation, and submissions. D. V. and I. M. led the experiment and contributed to the data analysis. S. E. provided theoretical support based on canonical-basis time-dependent Hartree-Fock-Bogoliubov theory. L. A. A., H. A. F., G. B., V. B., M. C., I. D., A. G., K. H., N. J., A. K., and T. M. contributed to the experiment. P. B., A. B., M. C., F. C. L. C., M. K., A. M., V. N., and O. S. contributed to the construction of the $\gamma$-ray spectrometer PARIS. All authors reviewed and commented on the manuscript. This article results from the Ph.D. thesis work of R. L.

%\section*{Competing interests}
%The authors declare no competing interests.

%\section*{Additional information}

%\textbf{Correspondence and requests for materials} should be addressed to R. L.

%\textbf{Peer review information} Nature thanks anonymous reviewers for their contribution to the peer review of this work.

%\textbf{Reprints and permissions} information is available at https://www.nature.com/reprints.

\end{document}